\documentclass[acmsmall,screen]{acmart}

\usepackage{mathptmx}
\usepackage[T1]{fontenc}
\usepackage[english]{babel}
\usepackage[utf8x]{inputenc}
\usepackage{graphicx}
\usepackage{setspace}
\usepackage{float}
\usepackage{tabularx}
    \newcolumntype{L}{>{\raggedright\arraybackslash}X}
\usepackage{booktabs}
\usepackage{subcaption}
\usepackage{wrapfig}
\usepackage{multirow}

\usepackage{xcolor}

\usepackage{hyperref}
\usepackage{amsmath,amsfonts}

\AtBeginDocument{%
  \providecommand\BibTeX{{%
    \normalfont B\kern-0.5em{\scshape i\kern-0.25em b}\kern-0.8em\TeX}}}

\setcopyright{none}
\renewcommand\footnotetextcopyrightpermission[1]{}

\begin{document}

\title{Investigating the Security of EV Charging Mobile Applications As an Attack Surface}


\author{K\lowercase{haled}~S\lowercase{arieddine}}
\email{khaled.sarieddine@mail.concordia.ca}
\affiliation{%
  \institution{The Security Research Centre, Concordia University}
  \city{Montreal}
  \state{Quebec}
  \country{Canada}
}

\author{M\lowercase{ohammad}~A\lowercase{li}~S\lowercase{ayed}}
\affiliation{%
  \institution{The Security Research Centre, Concordia University}
  \city{Montreal}
  \state{Quebec}
  \country{Canada}
}
\author{S\lowercase{adegh}~T\lowercase{orabi}}
\email{storabi@gmu.edu}
\affiliation{%
  \institution{Center for Secure Information Systems, George Mason University}
  \city{Fairfax}
  \state{Virginia}
  \country{USA}
}

\author{R\lowercase{ibal}~A\lowercase{tallah}}
\affiliation{%
  \institution{Hydro-Quebec Research Institute}
  \city{Montreal}
  \state{Quebec}
  \country{Canada}
}

\author{C\lowercase{hadi}~A\lowercase{ssi}}
\email{assi@encs.concordia.ca}
\affiliation{%
  \institution{The Security Research Centre, Concordia University}
  \city{Montreal}
  \state{Quebec}
  \country{Canada}
}

\renewcommand{\shortauthors}{Sarieddine et al.}

\begin{abstract} 
    The adoption rate of EVs has witnessed a significant increase in recent years driven by multiple factors, chief among which is the increased flexibility and ease of access to charging infrastructure. To improve user experience \textcolor{black}{and} increase system flexibility, mobile applications have been incorporated into the EV charging ecosystem. EV charging mobile applications allow consumers to remotely trigger actions on charging stations and use functionalities such as start/stop charging sessions, pay for usage, and locate charging stations, to name a few. In this paper, we study the security posture of the EV charging ecosystem against \textcolor{black}{a new type of remote} which exploit\textcolor{black}{s vulnerabilities in} the EV charging mobile applications as an attack surface. We leverage a combination of static and dynamic analysis techniques to analyze the security of widely used EV charging mobile applications. Our analysis \textcolor{black}{was performed on 31 of the most widely used} mobile applications including their interactions with various components such as the cloud management systems. \textcolor{black}{The attack, scenarios that exploit these vulnerabilities were verified on a real-time co-simulation test bed. Our discoveries indicate the lack of user/vehicle verification and improper authorization for critical functions, which allow adversaries to remotely hijack charging sessions and launch attacks against the connected critical infrastructure}. \textcolor{black}{The attacks were demonstrated using the EVCS mobile applications showing the feasibility and the applicability of our attacks.} Indeed, we discuss specific remote attack scenarios and their impact on EV users. More importantly, our analysis results demonstrate the feasibility of leveraging existing vulnerabilities across various EV charging mobile applications to perform wide-scale coordinated remote charging/discharging attacks against the connected critical infrastructure (e.g., power grid), with significant economical and operational implications. Finally, we propose \textcolor{black}{countermeasures} to secure the infrastructure and impede adversaries from performing reconnaissance and launching remote attacks using compromised accounts.
\end{abstract}

\maketitle

\section{Introduction}
\label{introduction}

Climate change and increased greenhouse gas emissions \textcolor{black}{are fueling society's} embrace of \textcolor{black}{a} green technology mindset. \textcolor{black}{Governments are diligently working on} shifting \textcolor{black}{the} traditional transportation system to a smarter one by utilizing Electric Vehicles (EVs). Many countries have already implemented policies to reach carbon neutrality, partly by adopting EVs with an aim to reach $30$\% EV market share by 2030. For instance, China's plan to reach carbon neutrality resulted in a significant $141\%$ increase in the deliveries of EVs in October 2021 \cite{regan_2020}. Similarly, multiple federal/provincial governments in North America along with various European countries have set policies to increase EV adoption in the next decade while banning the sales of new gasoline-powered passenger cars in the near future \cite{gyulai_2020, riley_2021}. Consequently, there is a rapid deployment of EV Charging Stations (EVCSs) to match EV adoption. Furthermore, governments are actively investing \textcolor{black}{in the EV charging infrastructure} to address the \textcolor{black}{lag of EVCSs} with respect to the increasing demand caused by the shift towards EVs \cite{canada_2021}.

The leading automotive companies are investing in incorporating advanced technological features in EVs and the Cyber-Physical System (CPS) associated with them \textcolor{black}{for} remote management. The cloud management system (CMS) \textcolor{black}{manages all operations and functionalities} of the public EVCS. As a result of the need to commercialize the EV charging ecosystem and to drive the adoption of EV technologies, the EV charging mobile application established itself as a core component of the EV ecosystem \cite{acharya2020cybersecurity}. Mobile applications are used by EV consumers to remotely control EVCS and manage charging operations (e.g., start and stop charging sessions) \textcolor{black}{through the CMS}. They also provide various online functionalities such as checking availability, and handling payments, to name a few. The importance of studying this component in a fine-grained manner lies \textcolor{black}{in} its ability to control the EVCSs and its wide distribution among customers. The existence of multiple vendors, along with minimal efforts to standardize the development and deployment of EVCS components has \textcolor{black}{exposed} the ecosystem and the interconnected critical infrastructure (e.g., power grid) to a wide range of remote attacks \cite{mohammad_paper, tony_paper}. Moreover, the massive adoption of EVs caused a compelling change in the transportation system and the power grid, not only increasing the \textcolor{black}{its} load but also increasing the complexity of the ecosystem as a whole \cite{akhras2020securing}. This adoption led to the rapid deployment of Internet-enabled EVCSs along with various supporting EV charging mobile applications. Moreover, the number of product and cross-product applications is increasing tremendously where one mobile application might have access to more than one EVCS network (e.g., Flo, Blink, and EVgo), thus, amplifying the impact by exposing more products and users simultaneously. It is worth highlighting that such extended remote functionalities provide adversaries with a new attack surface, which could be utilized to compromise \textcolor{black}{the} vulnerable mobile application and thus, exploit the underlying \textcolor{black}{EVCSs and their operation}.

A number of studies focused on the security of the EV ecosystem by exploring the security of the firmware, the installed management systems, and the communication link \cite{alcaraz2017ocpp, rubio2018addressing, elhussini2021tale, antoun2020detailed, tony_paper}. 
For instance, Nasr et al.~\cite{tony_paper} studied the security of the EVCS firmware and management systems and discovered 13 severe vulnerabilities. Furthermore, Antoun et al.~\cite{antoun2020detailed} presented a detailed security analysis of the ecosystem while assuming a highly privileged attacker that can fake, intercept, inject, and modify messages. Whereas Alcaraz et al.~\cite{alcaraz2017ocpp} studied the security posture of the OCPP protocol, which is the main protocol used to control EVCSs\textcolor{black}{,} and discovered its susceptibility to Man-In-the-Middle attacks. Despite such \textcolor{black}{efforts} to explore the security of various components within the EV charging ecosystem, there is a lack of knowledge about the security posture of the existing EV \textcolor{black}{charging mobile applications as an attacker} entry point. Moreover, there is a lack of understanding about the extended attacker capabilities and attack implications when leveraging vulnerabilities across widely used mobile applications to perform large-scale coordinated attacks against various stakeholders and components within the EV charging ecosystem.

In this study\textcolor{black}{,} we focus on investigating the security posture of the EV charging mobile applications as an attack surface on the EV charging ecosystem and the underlying infrastructure. \textcolor{black}{We are the first to systematically analyze the most widely used EV charging mobile applications that are utilized to manage EVCSs across various geographical locations. We focus on identifying design flaws in the interaction between the different components as it requires a deeper understanding of the ecosystem and advanced threat modeling.} We utilize reverse engineering and code analysis techniques to get insight into the functionalities instilled \textcolor{black}{in} the EV charging mobile applications, along with the security measures implemented by these \textcolor{black}{applications} (e.g., bot detection libraries). Consequently, guided by the static analysis, we perform dynamic functionality analysis to test the interactions of these applications with other components while identifying their main implemented EV charging-related functionalities that can be triggered remotely (e.g., remote start and stop). Furthermore, we analyze the traffic exchanged between the EV charging mobile application and the CMS to understand their interactions. \textcolor{black}{We also examined the exchanged} information by breaking SSL using Man-In-The-Middle (MITM) monitoring to unravel encrypted interactions. \textcolor{black}{Additionally, we utilize a deep understanding of the ecosystem and the available literature} to complement our analysis of the EVCS-Cloud communication \cite{tony_paper, ocpp, alcaraz2017ocpp, baker2019losing}. 

Our analysis of the exchanged traffic/interactions \textcolor{black}{allowed us to infer} state transitions between the mobile application and other entities in the EV charging ecosystem. We found that the analyzed mobile applications lack adequate EV ownership verification. Moreover, most applications implement improper authorization for initiating critical functions (e.g., start charging), which only binds users with EVCSs without binding/authenticating users to the EVs. This is closely related to the lack of adequate EV ownership, which allows insecure initiation of a critical function such as EV charging/discharging. 
\textcolor{black}{Indeed, we found a lack of proper ownership checking and improper authorization for critical functions (start and stop charging)}. While such vulnerabilities demonstrate the insecurity of the EV charging ecosystem, it also highlights the immaturity of the deployed software components, which hinders the advancement to reach the goals set by the industry. Furthermore, 29 out of 31 studied mobile platforms are susceptible to remote charging session hijacking while enabling mass (dis)charging attacks. Additionally, our analysis illustrates that 19 of the vulnerable EV charging mobile application platforms can be exploited to perform large-scale oscillatory load attacks on the connected infrastructure through \textcolor{black}{the} unauthorized remote start/stop charging capabilities. Consequently, we study the feasibility of synchronized remote charging attacks and their impact on the underlying infrastructure by studying transmission losses, \textcolor{black}{and} generation costs, along with overloading and transmission line tripping and power grid stability. Finally, we provide recommendations to prevent remote attacks utilizing the EV charging mobile application. To this end, we frame the main contributions of this work as follows:

\begin{itemize}
    \item \textcolor{black}{To the best of our knowledge, we are the first to study the security posture of EV charging mobile applications as an attack surface against the power grid. We study the interactions of the different components and validate them on a real-time co-simulation test bed.} Our findings demonstrate the insecurity of such mobile applications while highlighting design and implementation weaknesses that can be exploited to perform unauthorized operations on the underlying EVCSs. 
    
    \item We leverage static and dynamic analysis techniques to analyze the implementation of EV charging mobile applications and their interactions with various components within the ecosystem. We utilize Finite State Machines to provide an abstract representation and model the interactions of the mobile applications with the cloud management system and the EVCS counterparts. Our analysis indicates several vulnerable interactions due to unverified user/vehicle ownership, and improper authorization for critical functions. \textcolor{black}{We demonstrate such attacks by showing successful proof of concept attack scenarios that exploit the design flaws we discovered.}
   
    \item Given the feasibility of the identified vulnerabilities across widely used EV charging mobile applications, we investigate the implications of wide-scale remote attack scenarios by constructing synchronized botnet attacks that utilize the mobile applications as a recon to perform various unauthorized charging operations including large-scale voltage/frequency instability attacks on the power grid. We discuss practical attack implications against the stakeholders, precisely, and the connected power grid infrastructure. Finally, we propose design/implementation countermeasures to mitigate such attacks in the future.
\end{itemize}

The remainder of this paper is organized as follows. In Section \ref{sec:Background}, we present background information and basic concepts related to the EV ecosystem. In Section \ref{sec:methodolog}, we discuss the analysis methodology. In Section \ref{sec:results}, we discuss our findings in terms of identified interactions and vulnerabilities, along with attack feasibility. We discuss detailed attack implications against various stakeholders in Section \ref{sec:implications}. In Section \ref{sec:countermeasures}, we discuss a mitigation framework along with security measures that will help defend against such exploitation before concluding the paper in Section \ref{sec:conclusion}.

\section{System Model, Related Works, and Threat Model}
\label{sec:Background}

The EV charging ecosystem is a cyber-physical system, composed of interacting hardware and software components. In what follows, we provide details about these components and their interactions.

\begin{figure}[t]
	\centering
	\includegraphics[width=0.7\linewidth, height=6cm]{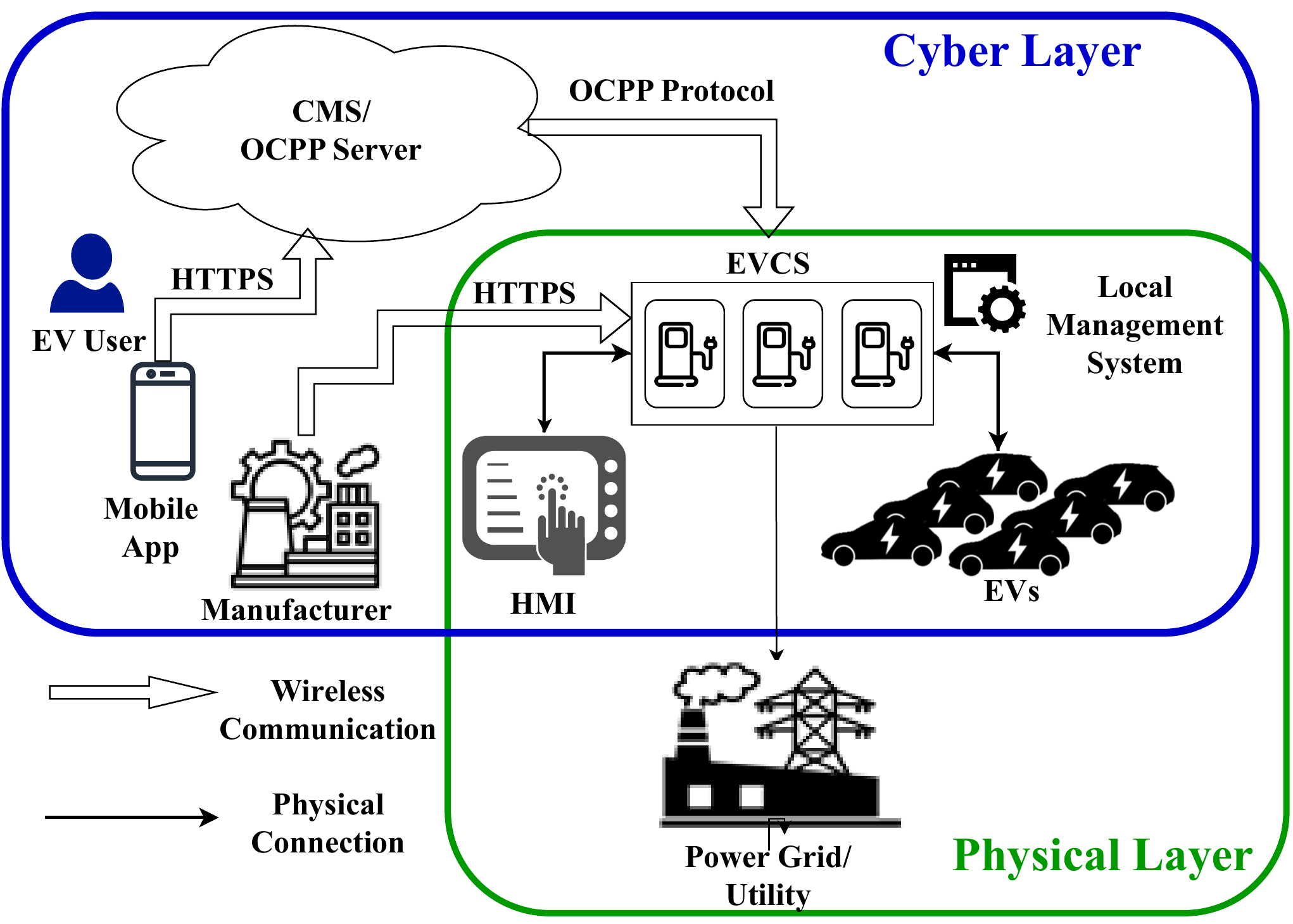}
	\caption{Overview of the EV charging ecosystem and its interactions. }
	\label{fig:ecosystemOverview}
\end{figure}

\subsection{\textcolor{black}{System Model}}
\textcolor{black}{The EVCS ecosystem incorporates multiple entities that collaborate and interact to provide a vital service to the customers (individuals and businesses). It is the main enabler for EVs that have been spreading rapidly due to governmental policies that have driven their adoption. The EVCS ecosystem consists of a cyber and a physical layer, as shown in Figure \ref{fig:ecosystemOverview}.}

\subsubsection{Cyber-layer}
\textcolor{black}{The Cyber Layer is composed of multiple software components coupled with the hardware/physical counterpart. The mobile applications are publicly available and distributed through application stores (Google Play and Play Store) These applications are needed by users to control EVCSs remotely and view EVCSs' status through their communication with the CMS. The mobile applications could either be operator-specific (manage EVCSs belonging to one operator) or multi-operator (manage EVCSs of multiple operators). The multi-operator mobile applications were introduced to simplify the charging process and enable EV roaming among different operators without the need for operator-specific subscriptions. Consequently, based on the above distinction the operator-specific mobile application communicates with the operator's CMS, whereas the multi-operator mobile applications communicate with the back-end of the application's owner which in turn forwards the requests to the respective operator's CMS using the Open Charge Point Interface (OCPI) \cite{ocpp}.}

\textcolor{black}{The CMS plays an equally important role in the ecosystem since it provides API endpoints for the mobile application to communicate with the EVCS. Each operator has their own CMS that is responsible for reservation, scheduling, payments, management, monitoring, etc. The CMS is the most computationally capable component and it is considered the main driver of the ecosystem. However, to control EVCSs, the CMS communicates with the EVCS using the Open Charge Point Protocol (OCPP).}

\textcolor{black}{The OCPP protocol is the de facto standard that is utilized to manage EVCS remotely. The OCPP defines two main roles, a lightweight client (EVCS) and a central server (CMS), which utilizes full-duplex communication over a TCP connection. The communication of the OCPP protocol is in the form of transaction functional blocks, where each entity requires a response to the initiated transaction. This standard is maintained and developed by an alliance of multiple companies working in the industry. Moreover, a connection is usually maintained between the EVCS and the original manufacturer which helps in collecting logging information about the performance of the station. We have validated these interactions between the different components and the responsibility of each entity, with our industrial partner Hydro-Quebec, a major North American utility.}
 
\subsubsection{\textcolor{black}{Physical-layer}} \textcolor{black}{The Physical Layer is represented by different entities. Namely, the EVCS hardware includes the human-machine interface that is used by the users to interact physically with the EVCS. After an EVCS is manufactured and bought by an operator, the manufacturer maintains a connection to push firmware updates remotely or can make the updates available online for the operator to manage the process. Moreover, the EVs have multiple hardware and software components including remotely accessible components such as an On-board Diagnostic Port and a CAN bus.} The EV charging ecosystem was established to match the demand of EVs and their need to charge. Two types of EVs are dependent on the EV charging ecosystem, which are the main foci when studying the security of the EV charging ecosystem: Plug-in Hybrid EVs (PHEVs) and Battery EVs \cite{wong2006battery, elhussini2021tale}. Other types of EVs such as Hybrid Electric Vehicles and Fuel Cell Electric Vehicles do not require external charging \cite{USDepEnergy}. Consequently, EVs connect to the charging stations using various standards (SAE J-1772/J-2293/J-2847/J-2836, IEC62196/61851, ISO/IEC 15118, and chAdeMO) \cite{elhussini2021tale, tony_paper, mohammad_paper} which are a part of the efforts to standardize communication in the EV charging ecosystem. 

\textcolor{black}{Moreover, }there are several EVCS classifications. Level 1 chargers (slow chargers) are being replaced by Level 2 chargers, which are mostly used commercially as public EVCSs. Moreover, Level 3 chargers (providing a higher charging rate) are being introduced to improve the user experience and decrease charging times \cite{mohammad_paper}. In this study, we focus on public EVCSs deployed by companies, governmental entities (e.g., Circuit Electric and ChargePoint), or private EVCSs that are made publicly available by the owner to earn extra income. It is worth highlighting that the EVCSs is also connected to the power grid (critical infrastructure) to draw the needed power for EVs to charge. 

\subsection{Overview of EV Charging Mobile Application Platforms}
\label{sec:overviewMobPla}
To manage the ever-increasing number of EVCS from different vendors, many companies have put in place a mobile application management platform for EV owners to monitor and control EVCSs. The mobile application is an indispensable component for the operation of the EV charging ecosystem. Consequently, we give an overview of the mobile application which encompasses its role and the communication protocols used when communicating with the different entities (CMS and EVCS). Mobile applications communicate with the CMS using Hypertext Transfer Protocol Secure  (HTTPS), which incorporates the Secure Socket Layer (SSL)/Transport Layer Security (TLS), to secure and encrypt the HTTP communication and to preserve the privacy of users by protecting against on-path attacks. Various functionalities are instilled in the mobile application that provides users with an interface to remotely monitor and control EVCSs as illustrated in Figure \ref{fig:ecosystemOverview}. To name a few, a discovery service is provided by mobile applications to allow users to find nearby EVCSs. Furthermore, a control service is instilled to\textcolor{black}{ allow remote start and stop of charging sessions and schedule charging}. 

Various vendors provide mobile applications that allow remote monitoring/charging. To name a few, ChargePoint, a leading operator, has over $500,000$ users. Similarly, ChargeHub is a popular mobile application that provides users with services to remotely control EVCSs. Indeed, the number of downloads of EV charging applications, which can be extracted from Google Play, shows the growth of the EV ecosystem, and the wide distribution of the mobile application among consumers. This sheds the light on the popularity and wide distribution of these applications that are used by many users. Moreover, the EVCSs and their management system in most cases belong to one or more different networks, thus allowing cross-application control. For example, the ChargePoint mobile application controls Blink, SemaCharge, and of course its own network. This highlights the importance of developing secure mobile applications due to its cross-operator collaboration which adds more heterogeneity to the ecosystem, especially with the lack of standardization.
\subsection{\textcolor{black}{Related Work}}
\label{sec:related}
\textcolor{black}{In this section, we survey and discuss previous work that tackled the security of the EV charging ecosystem's components. The security was analyzed from various perspectives, one of which discussed the security software component and the communication protocols, and the implications of the security vulnerabilities on the infrastructure. }

\subsubsection{Software Components Security} \textcolor{black}{Nasr et al. \cite{tony_paper}, studied and examined the security posture of the EVCS and their management systems. They managed to find vulnerabilities across 13 severe vulnerability classes in firmware and management systems (mobile applications and websites). It is worth mentioning that in \cite{tony_paper}, mobile applications were analyzed using only static analysis, whereas our analysis utilizes both aspects to understand the interaction of the different components without taking into consideration the interactions of the components and design flaws. Moreover, outside of academia Kaspersky Lab's team \cite{kaspersky} analyzed the security of ChargePoint home charging station and found significant vulnerabilities in its firmware and mobile management application. In our study, we provide a systematic and detailed analysis of the top 31 EV charging mobile applications. We are the first to address the lack of a comprehensive analysis of the EVCS mobile application that plays an integral role in the EVCS ecosystem. It is worth noting that the mobile applications used to manage EVCS are widely distributed and easily accessible which simplifies and makes it easier for the attacker to acquire, analyze, and compromise, compared to firmware or cloud management systems.}

\textcolor{black}{Moreover, the communication links have been also studied in the literature. Alcaraz et al. \cite{alcaraz2017ocpp} examined the communication protocol and presented a vulnerability in the OCPP protocol that allows for man-in-the-middle (MitM) attacks, thus interfering in the communication between the EVCS and the EV resource reservation service. Moreover, in \cite{occpdef} they address the vulnerabilities presented in \cite{alcaraz2017ocpp} and provide countermeasures. The security of the OCPP protocol has been improved, various security measures have been implemented to harden the communication protocol such as the addition of secure firmware updates, security logging, and event notification and security profiles for authentication (key management for client-side certificates) and secure communication (TLS) \cite{ocpp}. }

\subsubsection{Infrastructure}
\textcolor{black}{Mohammad et al. \cite{mohammad_paper} demonstrated the impact of compromising EVCSs on the power grid and launched attacks against it. Then discussed the non-linear nature of the EV charging load and simulated multiple attacks that can be launched against the power grid using these EVs. While the grid was able to recover after a 48 MW attack utilizing traditional residential loads, \textcolor{black}{a smaller 30 MW EV load attack is able to completely destabilize the grid.} Moreover, in \cite{khan2019impact}, the authors study how a botnet of compromised EVs and fast-charging direct current stations can be utilized to launch cyber attacks on the power grid and its implications on the transmission and distribution networks. Moreover, in \cite{clement2009impact, leemput2014impact, dubey2015electric, morais2014evaluation, shafiee2013investigating} they study the implications of EV charging on the grid and discuss some mitigation techniques. Moreover, in \cite{soykan2021disrupting} they discuss the use of SMS phishing as a social-based attack. Where an attacker can send spoofed text messages to the users advertising a discount (20\% off when the users charge their vehicle at noon). Consequently, they studied the impact of such an attack on the grid. However, mobile phishing attacks require knowledge of the mobile phone numbers of EV users in a certain target area, which affects the feasibility of acquiring such information. Furthermore, in \cite{soykan2021disrupting} the attack depends on the susceptibility of users to the demand and response phishing attack. Whereas, in our study, we demonstrate how and when can an attacker get information to perform an orchestrated attack that might impact the stakeholders and study the feasibility of acquiring such information. Moreover, we demonstrate vulnerabilities that allow us to exploit the communication between the mobile application and the charging station. The discovered design flaws allow the adversary to charge any vehicle connected to an EVCS which could be used later in load-altering attacks.}

\subsubsection{Difference Over Related Work}
\textcolor{black}{Zhou et. al. \cite{zhou2019discovering}, studied home IoT communication with the cloud to detect device interaction issues. They performed traffic analysis by intercepting mobile application traffic with the cloud, and the IoT devices' interaction with the cloud. However, due to the limited access to the public EVCS infrastructure, we use functional analysis to monitor the EVCS state changes based on the mobile applications' user behavior. Moreover, we record user input in the application (e.g., user information, credit card, vehicle information) to understand the semantics of the traffic being sent. Utilizing traffic analysis similar to \cite{zhou2019discovering}, the origin and the meaning of the traffic being sent would be lacking. The adversary in \cite{zhou2019discovering} needs to have access to the charging infrastructure to detect vulnerabilities, however, in our analysis, we only utilize the mobile application as a means to get information about the whole ecosystem. Furthermore, in \cite{tony_paper}, the authors focused on static analysis to analyze the vulnerabilities in mobile applications, which limits their ability to analyze the interactions of the ecosystem while utilizing the mobile application as an attack vector. To the best of our knowledge, the analysis of the EVCS mobile application and its interaction with other components (CMS and EVCS) has not been done before}.

\subsection{Threat Model}
\label{sec:threatmodel}
We consider a remote adversary with access to one or more mobile applications distributed on the Google Play Store and Apple store. Moreover, we consider the remote adversary is able to create an account on these mobile applications. Similar to \cite{zhou2019discovering}, we do not assume any forms of software bugs or protocol vulnerabilities. The adversary relies on understanding the interactions between the components by utilizing various analysis methods to identify vulnerabilities and understand the interactions between the mobile application and the CMS. The analysis methods range from reverse engineering and white-box testing to functionality analysis, system fuzzing, and black-box analysis. The attacker aims to utilize mobile applications as an entry point to target EVCSs with connected vehicles. The adversary's goal is to exploit design flaws in the interactions among the different entities (e.g., EVCS, CMS, Mobile application) to hijack or initiate an unauthorized charging session remotely without compromising legitimate user accounts. Moreover, the adversary's ultimate goal is to leverage illegal charging sessions to perform large-scale, coordinated botnet attacks against the underlying critical infrastructure (e.g., the power grid) and the EV users.

\section{Methodology}
\label{sec:methodolog}
\begin{figure}[t]
    \centering
    \includegraphics[width=0.8\linewidth]{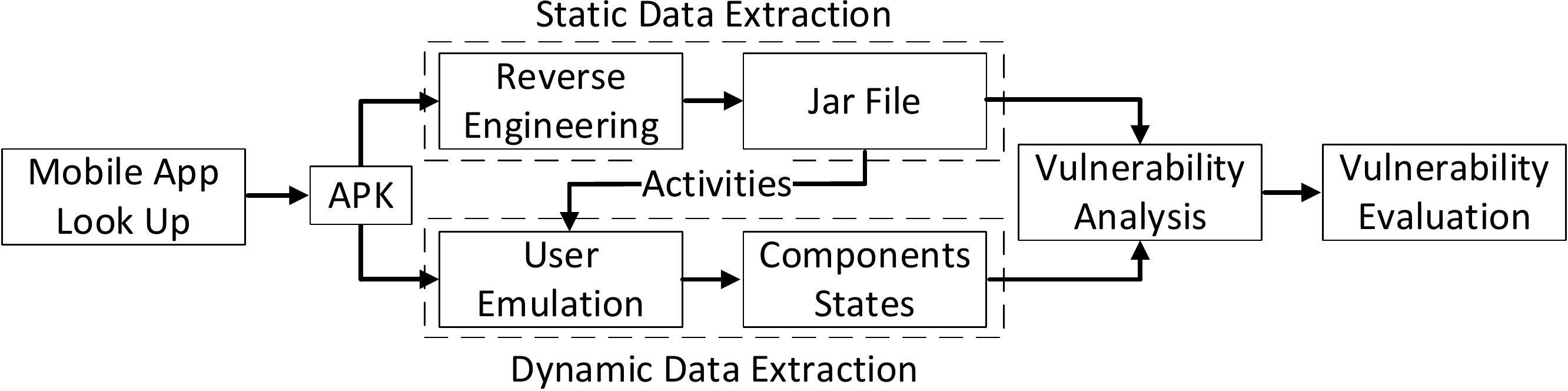}
    \caption{The overall mobile application lookup and vulnerability analysis methodology.}
    \label{fig:methodology}
\end{figure}

In this section, we elaborate on the analysis methodology to identify vulnerabilities that allow adversarial accounts to control charging sessions for vehicles they do not own. As shown in Figure~\ref{fig:methodology}, we combine static (reverse engineering and code review) and dynamic analysis techniques (functionality and traffic) to perform vulnerability analysis and assessment \textcolor{black}{of} the identified mobile applications. 
First, we start by fetching EV charging mobile applications, then for each product, we extract data for analysis by applying reverse engineering techniques on their binaries. Second, we extract network traffic during the functionality analysis while emulating \textcolor{black}{the} user behavior of the application; consequently, we analyze the application states and behavioral changes to abstract the system interactions \textcolor{black}{and} then evaluate it to find flaws. We provide details on the proposed methodology components below.

\subsection{Mobile Application Look Up}
In this section, we \textcolor{black}{discuss} our selection strategy for mobile applications. According to Statista \cite{dea_29_2021}, Android maintained its position as the leading mobile operating system for mobile phones \textcolor{black}{with} about $73\%$ \textcolor{black}{market} share. We look for EV charging mobile applications on the Google Play store, which is the main platform used by Android users to download applications. Furthermore, \textcolor{black}{we automatically fetch 50 mobile applications from the Google play store. Then, we choose the mobile applications and filter them based on the features they possess by automatically searching the description for keywords such as start/stop charging. After further analysis, we discarded 8 mobile applications that are either EV charging calculators or not related. Our analysis focuses on mobile applications that provide remote control functionalities to control public EVCSs, consequently, we analyze the applications manually to ensure the existence of these functionalities as they pose a real danger \textcolor{black}{to} the power grid when compromised at scale \cite{mohammad_paper}.}

Based on the prior differentiation between the applications according to their capabilities, we identify and classify the applications \textcolor{black}{into} three types, as shown in Table \ref{table:typesOfEVapps}. Type 1 are applications that allow users to have an overview of the ecosystem and control the charging session, while Type 2 applications are used to perform reconnaissance activities and can only show an overview of the system. \textcolor{black}{Whereas Type 3 mobile apps are developed to target home EVCS owners or businesses with private EVCSs, limiting its impact, especially from a power grid perspective. Type 3 apps could possess vulnerabilities, however, it is considered out of the scope of this work as we focus on mobile applications that provide access to public EVCSs thus, increasing the impact of an attack on the power grid. Namely, we focus on Type 1 applications \textcolor{black}{that} can be used to perform attacks on the power grid because of their special capability to control the EVCS and its operations. Additionally, Type 1 and 2 can be used by an adversary to prepare for their attacks by analyzing the availability of EVCSs and their usage trends, as discussed in Section}~\ref{sec:feasibility}. 

\textcolor{black}{Finally, we fetch the remaining mobile applications and download/install their \texttt{APKs}. It is crucial to highlight that any security concern discovered in the communication between the mobile application and the CMS applies to both Android applications and iOS since they rely on the same back-end that handles their requests in most cases. Therefore, we assume that our analysis methodology and results can be generalized to both platforms, respectively. However, confirming this will be considered for future work.}

\begin{table*}[t]
    \centering
    \scriptsize
    \caption{Types of EV charging mobile applications based on their abilities. $^{a}$ Indicates the mobile application operators that possess Flaw 1 (Unverified Ownership). $^{b}$ Indicates the mobile application operators that possess Flaw 2 (Improper authorization for a critical function). $^{c}$ Indicates the mobile application operators that possess Flaw 1 and 2 but mitigate them by requiring information only found physically on the EVCS HMI.}
    \begin{tabular}{p{3cm} c p{9cm}}\toprule
        \textbf{Description}         & \textbf{Type} & \textbf{Application Names}\\\midrule
        \multirow{2}{12em}{Remote control of charging sessions and system overview} & \multirow{2}{1em}{1} & 
        \textcolor{black}{\textbf{Remote Start Charging:} ChargeHub$^{{a, b}}$ - Electrify Canada$^{{c}}$ - PodPoint$^{{a, b}}$ - Electrify America$^{{a, b}}$ - EVDC$^{{a, b}}$ - Semma Connect$^{{a, b}}$ - eCharge Network$^{{a, b}}$ - Tata Power EZ Charge$^{{c}}$ - Flo EV Charging$^{{a, b}}$ - BC Hydro EV$^{{a, b}}$ - Circuit Électrique$^{{a, b}}$ - PlugShare$^{{a, b}}$}\\
        &&\textcolor{black}{\textbf{Remote Start and Stop Charging:} Petro-Canada EV$^{{a, b}}$ - ChargePoint$^{{a, b}}$ - vend-electric$^{{a, b}}$ - Anywhere Charging$^{{a, b}}$ - Electromaps$^{{a, b}}$ - Ionity$^{{a, b}}$ - Volta Charging$^{{a, b}}$ - Charge Assist$^{{a, b}}$ - Virta$^{{a, b}}$ - Global Charge$^{{a, b}}$ - EV Charging By NewMotion$^{{a, b}}$ - EV Match$^{{a, b}}$ - EcoFactor Network$^{{a, b}}$ - FastNed$^{{a, b}}$ - EVgo$^{{a, b}}$ - Greenlots$^{{a, b}}$ - EVduty$^{{a, b}}$ - EV Connect$^{{a, b}}$ - NextCharge$^{{a, b}}$}
        \\ \midrule
        System overview & 2 & Zap-Map - Charge Map - EVMap - Kazam EV - Chargeway - Charge Finder - Open Charge Map - EV Stations\\ \midrule
        Home charging optimization & 3 & EV Energy - OptiWatt - Monta EV Charging \\
        
        \bottomrule
    \end{tabular}
    \label{table:typesOfEVapps}
\end{table*}
 
\subsection{Static Analysis}
\label{sec:static}
We aim at documenting and understanding the functionality of mobile applications and \textcolor{black}{their} utilized libraries. We used static analysis to understand the security measures implemented by EV charging mobile application vendors to thwart automated bot attacks by examining libraries and artifacts found in the binary files, along with understanding the structure of the mobile applications (Figure~\ref{fig:static analysis}) to perform a systematic functionality analysis. Thus, we utilize reverse-engineering of the \texttt{APK}, which is an archive package that contains a manifest file with the package name, activity names, hardware features support, permission, and other configurations. The \texttt{APK} also contains the certificates for the application, a \texttt{lib} directory holding compiled libraries used by the application, and a file with compiled application code in the \texttt{dex} file format, which can be interpreted by the Dalvik Virtual Machine (DVM) and the Android run-time environment.

We extract all the files using apktool \cite{apktool}, which disassembles all the resources and extracts the application Manifest and the \texttt{dex} files. Consequently, we use the extracted dex files and convert them to a \texttt{JAR} file using dex2jar \cite{dex2jar} utility. \textcolor{black}{We then} input the file into jd-gui \cite{jdgui} to browse the underlying Java source code. We then analyze the extracted jar files using white-box analysis techniques to check the application resources (e.g., libraries and their functionality, certificate signing techniques used by the application developer, etc.). Further, we extract resources from the \textcolor{black}{generated reports} (e.g., the activities used in the mobile application and the flow of activities), which allow us to perform a detailed functionality analysis and identify libraries used in EV charging mobile applications by using MobSF \cite{mobsf} and LiteRadar \cite{literadar}. Consequently, we check the functionality of the libraries used (e.g., Google reCAPTCHA, hCAPTCHA, Anti-location Spoofing). \textcolor{black}{The understanding created by studying the information obtained through static analysis is used to systematically guide the following step which is the dynamic analysis.}

\subsection{Dynamic Analysis}
\label{sec:dynamic_analysis}
\begin{figure}
	\centering
	\begin{subfigure}[b]{0.48\textwidth}
		\centering
		\includegraphics[width=\textwidth]{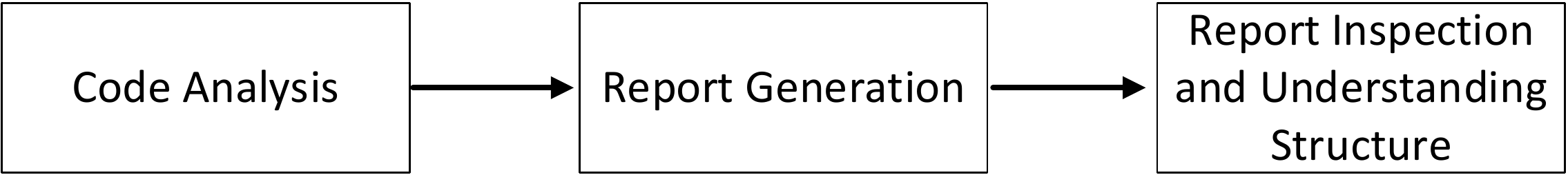}
		\caption{Static}
		\label{fig:static analysis}
	\end{subfigure}
	\hfill
	\begin{subfigure}[b]{0.48\textwidth}
		\centering
		\includegraphics[width=\textwidth]{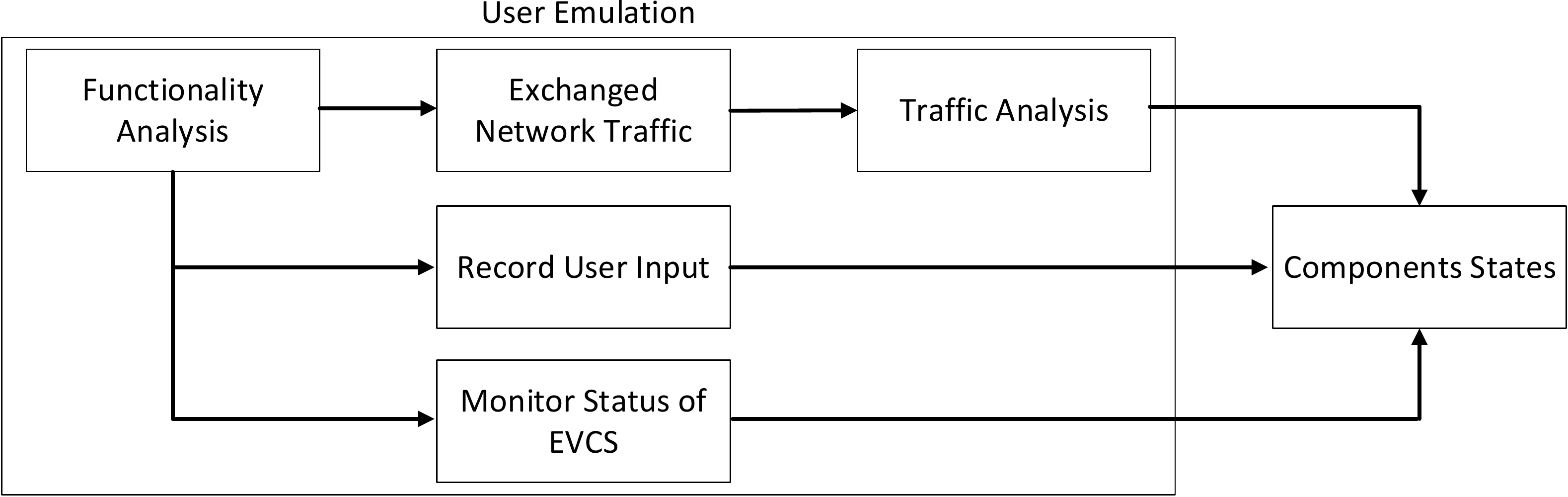}
		\caption{Dynamic}
		\label{fig:dynamic analysis}
	\end{subfigure}
	\caption{Overview of the static  and dynamic analysis methodologies.}
	\label{fig:method-dynamicandstatic}
\end{figure}

We rely on dynamic analysis (Figure~\ref{fig:dynamic analysis}) to complement our static analysis method and provide a holistic and comprehensive \textcolor{black}{assessment} of all the interactions and functionalities provided by the mobile application. The dynamic analysis \textcolor{black}{is performed through} functionality analysis, recording user input, traffic analysis, and monitoring system state changes. 

\textbf{Functionality Analysis:}
We perform functionality analysis to collect data and identify system states to understand the communication between different entities. Specifically, we seek to answer the question of ``How the adversary can utilize the interaction vulnerabilities and weaknesses to connect to a remote EVCS and control its operations?'' 
 
Guided by the static analysis that allowed us to understand the structure of each mobile application by mapping \textcolor{black}{its} activity flow, \textcolor{black}{we attempt} to systematically cover all the possible scenarios to systematically perform a detailed functionality analysis \cite{yang2015static, azim2013targeted}. We analyze each mobile application by manually mimicking regular users' behavior and operations and triggering every functionality possible in the mobile application. Throughout our analysis{,} we discover that some functionalities are strictly prohibited based on the location of the device. \textcolor{black}{For example}\textcolor{black}{,} Petro-Canada EV prevents initiating charging requests if the \textcolor{black}{users'} location is not \textcolor{black}{in the vicinity of} the EVCS. However, to mitigate that during our functionality analysis, we spoof the location of the device by broadcasting a location close to the EVCS. We utilized GPS JoyStick ADB Shell \cite{spoofLocation} to spoof the \textcolor{black}{device's location}. It is worth mentioning that other applications (e.g., Electric Circuit) notify the user that they are far away from the location of the charging station however, it does not restrict the user from initiating a charging request. Furthermore, some applications (e.g., electromaps) \textcolor{black}{detect} that the user is far from the EVCS even if the location was spoofed. This is attributed to the IP-Geolocation services used to detect the location of the originating IP, which can be circumvented using a VPN that routes our traffic through a private tunnel that \textcolor{black}{appears to originate} from the same country as the charging station.

\textbf{Traffic analysis:}
While emulating user behavior and performing user actions, we capture \textcolor{black}{the} traffic generated by the mobile application to understand the information that is being sent and the interactions between the mobile application and the CMS similar to \cite{zhou2019discovering}. We trigger important functionalities of the application such as sign up/in and start/stop charging. However, most applications use trusted Certificate Authorities (CA) to protect user privacy by encrypting the communication between the mobile application and the CMSs \cite{zhou2019discovering}. To decrypt the communication we utilize an un-rooted device with \textcolor{black}{the} Android 7 operating system 

Moreover, since Android OS with version ($\geq$ 7.0) does not trust \textcolor{black}{user-installed} certificates by design. \textcolor{black}{Thus, t}o run applications on an un-rooted device with \textcolor{black}{user-installed} certificates, we create a virtual space on the phone that allows \textcolor{black}{running} Android \texttt{APK}s as plugins by utilizing VirtualXposed \cite{says_2021}. Consequently, we perform API hooking to bypass certificate pinning/verification by \textcolor{black}{using} Inspeckage Package Inspector \cite{ac-pm}. Moreover, \textcolor{black}{in} some applications we bypass certificate verification
, by reverse engineering the application using APKTool \cite{apktool}, \textcolor{black}{followed by injecting} code into the application. 
We then repack and sign the application before installing it. Consequently, we utilize Burpsuite \cite{portswigger}, which operates as a proxy server between the mobile application and the target server to intercept, inspect, and modify raw traffic passing in both directions. Our analysis was not intrusive, however, it allowed us to unravel \textcolor{black}{and understand} the communication between the mobile application and the CMS. 

\section{Results}
\label{sec:results}
We utilize different methods to infer and analyze the interaction of the main components within the EV charging ecosystem during user and device registration and when initiating charging requests. \textcolor{black}{We extract the capabilities instilled in each mobile application. The capabilities recorded for each mobile application include remote start charging and remote stop charging. Moreover, we record the remote control restrictions that are implemented by the mobile application vendors to hinder \textcolor{black}{the} illegal or abnormal usage of a charging station. The vendors limit the flexibility for the user to initiate charging requests based on: (i) location proximity of the users (i.e., must be near the target charging station to initiate requests), (ii) IP Geolocation info (i.e., charging requests should originate from the same country/area as the target charging station), and (iii) user entered charging station ID that is found on the target device.}

Our preliminary analysis shows that all applications provide their users with a remote start charging service. Moreover, only 13 (e.g., Flo EV Charging, Plugshare, etc.) mobile applications do not provide \textcolor{black}{stop-charging} functionality forcing users to remove their cars when they finish charging. It is important to note that only two applications (Petro Canada EV and Electromaps) check the integrity of the users by validating the location of the device, whereas only one application (Electromaps) checks the locations of the originating IP of the charging request. Finally, two applications force the user to input a station ID or scan a QR code to initiate charging (Tata Power EZ Charge and Electrify Canada).

\subsection{Inferred Interactions}
As described in Section~\ref{sec:dynamic_analysis}, we leveraged dynamic traffic analysis to capture and infer the interactions between the mobile application and the CMS. As an adversary, we consider the communication between the EVCS and the CMS as a black box. However, the communication between the other entities can be inferred from the different states that the mobile application goes through while performing different actions. Moreover, while previous work presented in \cite{tony_paper, ocpp, alcaraz2017ocpp,baker2019losing} complement our analysis of the communication between the EV, EVCS, and the CMS, they provide us with some additional insights about such communication and interactions. Specifically, it has been shown that the communication between the EV and the EVCS lacks proper security measures, which renders the underlying equipment vulnerable to remote attacks. For instance, Baker et al.~\cite{baker2019losing} were able to eavesdrop on the Pulse-Width Modulation (PWM) communication, which \textcolor{black}{is} utilized by IEC 61851 \cite{schmutzler2013evaluation, schmutzler2012distributed} for safety-related signaling mechanism between EVs and EVCSs. Furthermore, despite the added security features into the ISO/IEC 15118 (e.g., Signal-Level Attenuation Characterization and TLS encryption), the works in \cite{schmutzler2012distributed,schmutzler2013evaluation, baker2019losing} highlighted the improper deployment of these features in practice. Moreover, as highlighted in \cite{mohammad_paper}, most of these security features are optional and are commonly ignored by manufacturers, thus, rendering devices vulnerable in real life.

\textbf{User and EVCS Registration.} To this end, we analyzed the EVCS charging applications listed in Table \ref{table:typesOfEVapps} of Type 1 to infer the interaction between the different entities upon user registration, EVCS registration, and upon sending a charging request. We identify the main interactions with the CMS during \textcolor{black}{the} registration of a new user and an EV charging station (EVCS). During user registration, each user is assigned a unique identifier, which allows the user to log into the platform and use existing functionalities. There are several options for the user identifiers such as email/password combination or authentication tokens to name a few. The unique user identifier is transferred to the CMS upon user registration on a given platform and then used later for authentication purposes. On the other hand, when an EVCS is installed and made available for the public, the operator needs to register it with the CMS by sending/registering its unique identifier. This information gets saved in the CMS and used by the mobile application to \textcolor{black}{identify a charging station}.

\textbf{Initiating Charging Requests.} 
After the user connects the vehicle to the charging station using one of the available connectors, the user must initiate a charging request using the mobile application by selecting the desired charging station (e.g., using a map view). The application embeds the unique user identifier along with the selected station's ID in the message sent to the CMS, respectively. The CMS then sends a start charge request to the charging station with the respective ID/info. Consequently, the EVCS checks for any connected vehicle before initiating the charging session. Note that in case no vehicle was connected at the time when the user initiates a charging request, the EVCS will wait for a grace period (e.g., 5 minutes) to provide the EV owner with sufficient time to plug the charger \textcolor{black}{into} the vehicle. Otherwise, if a car was found to be connected to the EVCS, it will initiate the charging session by sending a confirmation message to the CMS, this is inferred by monitoring the changes on the mobile application user interface. Once the CMS receives the confirmation, a correlation between the user identifier and the EVCS identifier is established. Finally, the CMS relays \textcolor{black}{the} charging confirmation sent by the EVCS to the mobile application. 

\begin{figure}[t]

	\begin{subfigure}[b]{0.28\textwidth}
			\centering
			\includegraphics[width=\textwidth]{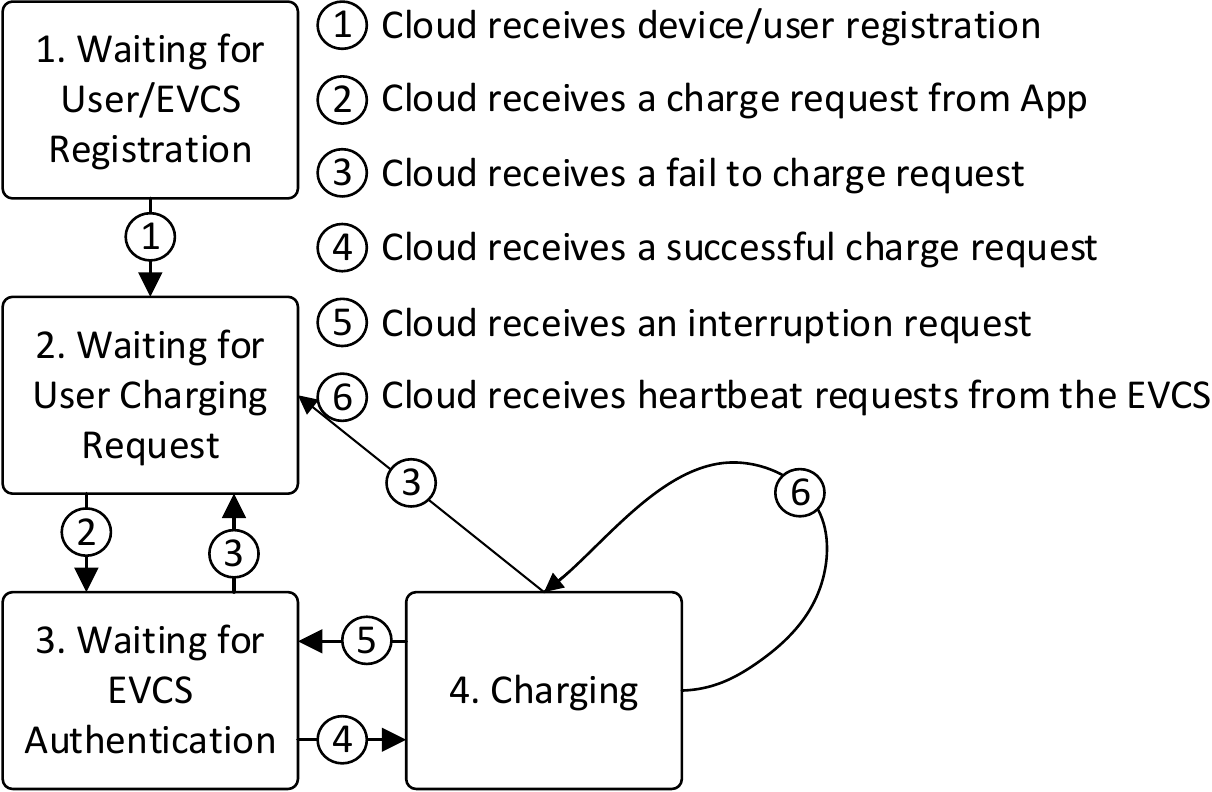}
			\caption{\scriptsize Cloud management system (CMS)}
			\label{fig:stateCMS}
	\end{subfigure}
	\hfill 	
	\begin{subfigure}[b]{0.35\textwidth}
			\centering
			\includegraphics[width=\textwidth]{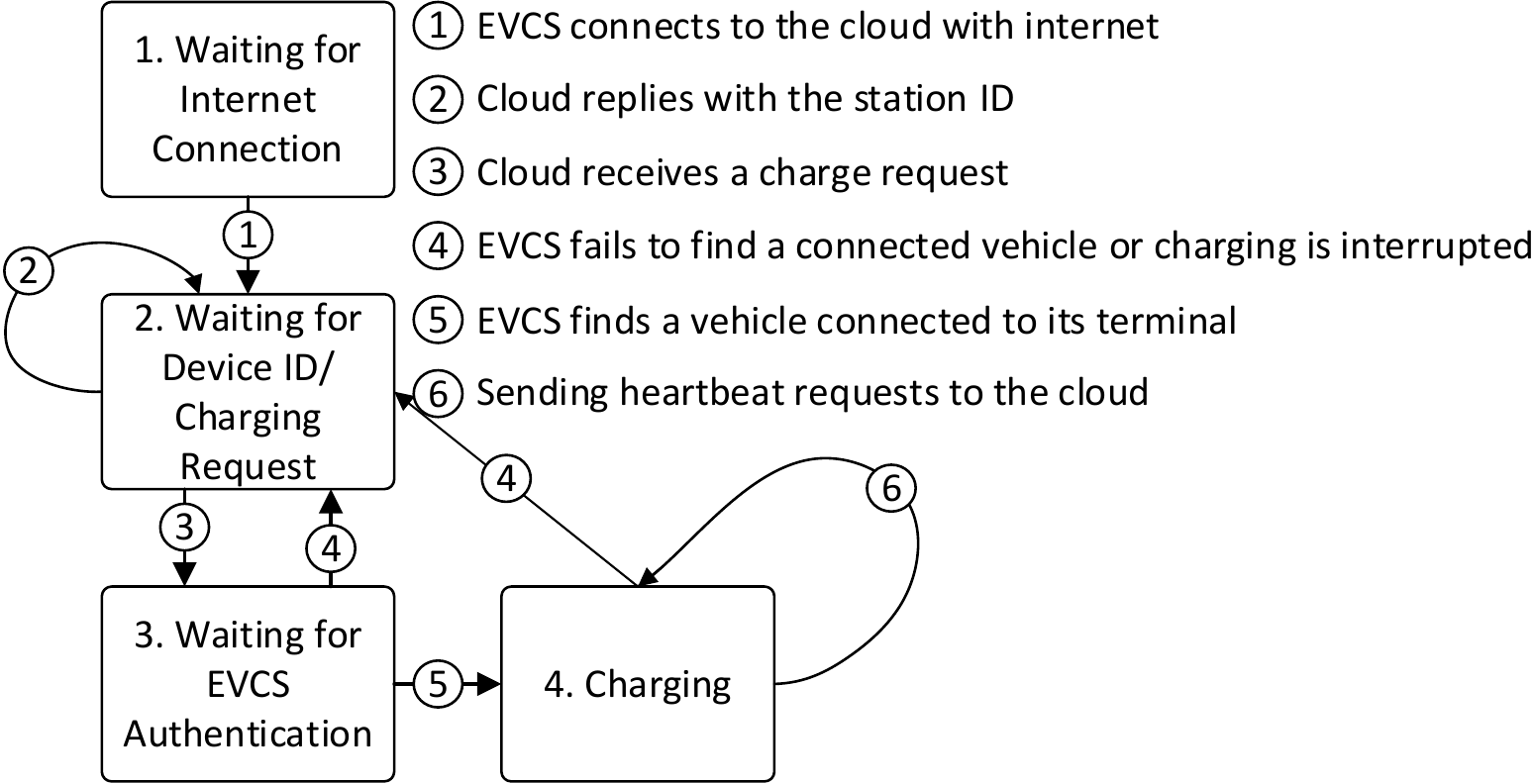}
			\caption{\scriptsize EV charging station (EVCS)}
			\label{fig:stateEVCS}
	\end{subfigure}
	\hfill	
	\begin{subfigure}[b]{0.32\textwidth}
			\centering
			\includegraphics[width=\textwidth]{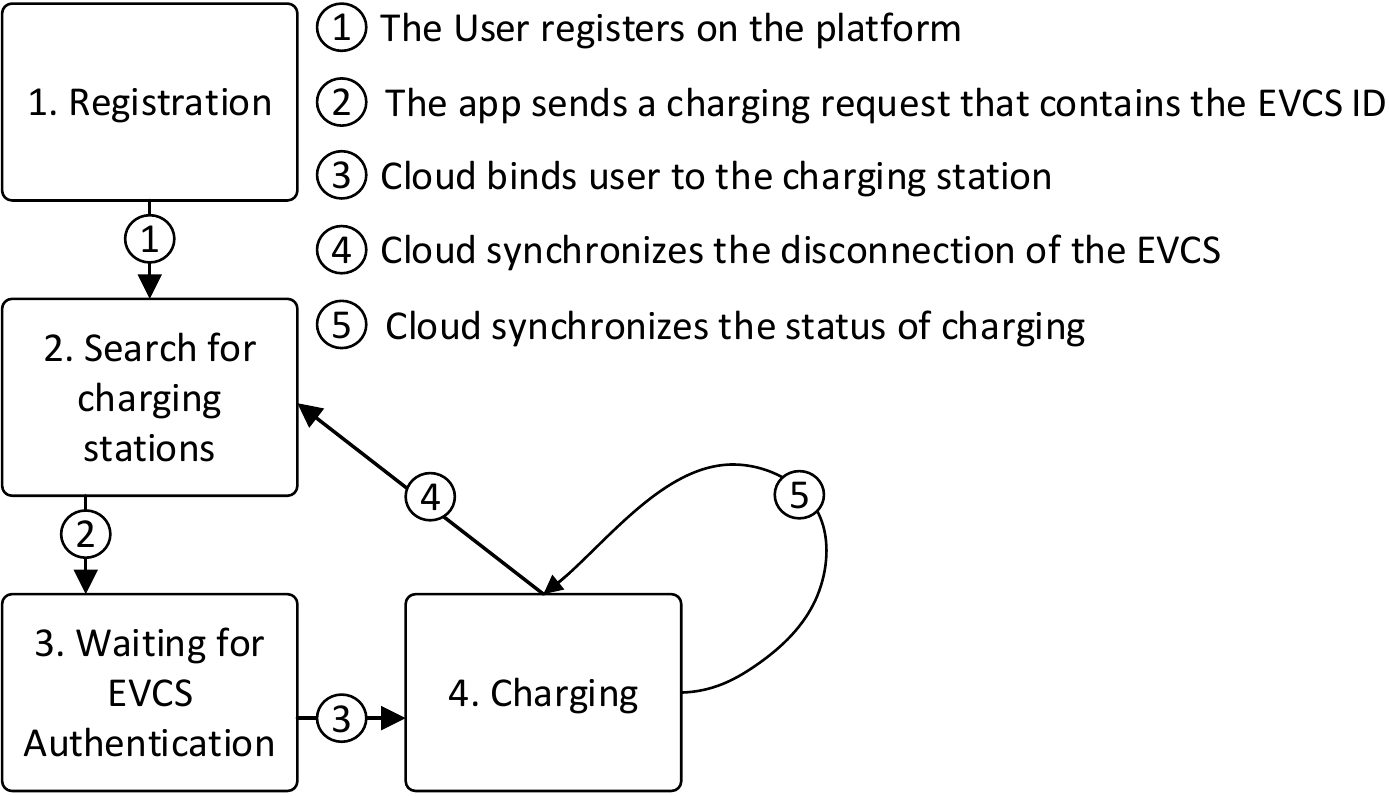}
			\caption{\scriptsize Mobile application}
			\label{fig:stateMobileApp}
	\end{subfigure}
	\caption{High-Level state machines for the three interacting components within the ecosystem.}
	\label{fig:states}
\end{figure}

    

To put this in a better context, we describe the state transitions inferred from the analysis of different platforms provided by the various EVCS operators in the industry. \textcolor{black}{We validated the interactions that helped us derive the states of the components on a real-time co-simulation testbed and two EVCSs acquired from one of the biggest North American operators. Any action triggered on the mobile application has a cascading effect on the other components and will alter the state of the other components. For example, whenever a user initiates a charging session from the mobile application} and transitions from S2 to S3 as illustrated in Figure \ref{fig:stateMobileApp}, the EVCS and the CMS will transition from S2 to S3 and eventually S4, the state of each component is dependent on the actions done by the other components. The different components making up the EVCS ecosystem are tightly coupled. The transition of one entity from one state to the other would change the current state of the ecosystem. An ideal system must strictly maintain the \textcolor{black}{three-entity} state machine. The legitimate states of the EVCS ecosystem are depicted as a 3-tuple combination. The CMS, EVCS\textcolor{black}{,} and the mobile application must strictly maintain the following 3-tuple states at all \textcolor{black}{times} to avoid potential attacks. \textcolor{black}{For example, if an attacker is able to induce the cloud and the EVCS to transition to state S3 while the legitimate user is still in state S2 shows that the charging session was hijacked by a third party (adversary). Moreover, another 3-tuple state if triggered can also lead to similar consequences by forcing the cloud and the EVCS to transition to state S4 whereas the legitimate user mobile application is in state S1 or S2. Whenever a legitimate user looking to charge his vehicle using a certain EVCS the cloud state and the EVCS state should allow future transitions rather than being blocked. If user B is charging user A is not allowed to access the EVCS remotely.} Note that all components have the same numeric state at all times, except at S1, where the CMS can be in S1 while the other components could be in either S1 or S2. \textcolor{black}{Consequently, through this work we aim at dissecting the intricate interactions of the components to trigger illegal states. The state machine shows the general operation of the EVCS ecosystem and shows the type of information that is shared during the process. Thus, through our analysis, we identify several vulnerabilities related to the trust model adopted in these ecosystems. Moreover, a deep understanding of a system allows us to identify flaws in the ecosystem design which are related to how the components interact with each other. Thus, we performed a systematic unraveling of the different components and their interactions.}

\subsection{Identified Vulnerabilities}
\label{sec:vuln}
In what follows, we present examples of the identified vulnerabilities that can be exploited to perform remote attacks against the EV charging ecosystem and the various involved stakeholders (e.g., EV consumer\textcolor{black}{s} and the power grid). We leveraged the described analysis methodology in Section~\ref{sec:methodolog} along with the inferred traffic/interactions in the previous sub-section to identify vulnerabilities.

Specifically, we discovered three major security weaknesses that are inherited from design and implementation flaws in the studied EV charging mobile application (Type 1). The EVCS charging platform \textcolor{black}{does} not strictly comply with the legitimate 3-tuple states. We found that the three entities stay in multiple unexpected 3-tuple states. The first unexpected state is (S4, S4, S1/S2); the CMS and the EVCS transition to the charging state, whereas the mobile application user is either still in registration or EVCS discovering state. This illegal state combination when exploited by an adversary could allow for remote charging/discharging session hijacking. 
	
In what follows, we elaborate further on the root cause of such behavior, the identified vulnerabilities\textcolor{black}{,} and their implications. It is worth noting that some mobile applications (e.g., EVMatch) mitigate the first unexpected state by forcing the user to reserve a spot \textcolor{black}{beforehand}. Whereas, other applications (e.g., Tata Power EZ Charge) hinder remote hijacking by forcing users to scan a QR code on the EVCS. However, when statically analyzing the mobile applications, QR codes are saved in a temporary file in the external SD card, which allows an on-device attacker to get access to that information to hijack the charging sessions. It is worth noting that some applications hide the access to charging behind payment gateway (e.g., buying store credit). However, adversaries can overcome this by buying store credit, which will provide access to the charging infrastructure. \textcolor{black}{In what follows we focus only on the flaws that can be used to manipulate the EVCSs for attacking the grid.}

\textbf{Flaw 1 (F1): Unverified ownership.}
Ideally, an EV user should be the owner or authorized user of the vehicle. Thus, the EV user should be the sole entity to authorize any form of control or action on the vehicle. Interestingly, our static and dynamic analysis results indicate that the mobile applications do not verify user ownership over target vehicles when initiating charging requests. In other words, an EV charging mobile application user can initiate a charging request to any vehicle connected to the network since both the mobile application \textcolor{black}{nor} the CMS do not have a mechanism to bind the application user to the target vehicle. Given that all communications of the mobile application go through the CMS, it is imperative to have the CMS verify critical operations such as vehicle identification and ownership management. On the other hand, access to vehicles from an unauthorized user is not verified within the EV charging mobile application platforms. Therefore, rendering the EV exposed to any user who can claim ownership and control over its charging functions when connected to the EVCS. Thus, leading to unexpected behaviors and potentially exploitable states.

\textbf{Flaw 2 (F2): Improper authorization for a critical function.}
Starting and stopping charging operations on a given EVCS are considered examples of critical functions, which could be abused by adversaries (unauthorized users) to destabilize the operations of the EVCS and the connected power grid. This was clearly demonstrated in \cite{mohammad_paper}, where the authors measured the impact of mass charging and stopping operations on the power grid. Ideally, an EV owner/operator should be the only user authorized to perform critical functions on the vehicle. Exposing critical functionalities essentially allows adversaries to control charging sessions, which is considered as the first step \textcolor{black}{toward} initiating mass charging attacks on the grid.  Nevertheless, our analysis results indicate that there is a lack of authorization, which allows any actor to perform critical functionalities on the \textcolor{black}{connected EVs}. This is closely related to our first finding (F1), it is mainly due to the fact that the binding step happens only based on the user and charging station IDs 
without further verification of the EV ownership or binding to specific mobile users. Therefore, an adversary can utilize fake accounts to hijack sessions.

\subsection{Attack Scenarios}
\label{sec:attackScenario}
An attacker can leverage the discussed vulnerabilities to launch various malicious activities against the EVCS and its operations such as remote charging sessions hijacking. To do this, attackers need to control a number of adversarial accounts (\textcolor{black}{i.e.}, bots), which provide access to existing charging services through mobile applications. Note that adversaries do not need to exploit or hijack user accounts to create the required botnet. \textcolor{black}{The attackers can easily create their own botnet of legitamate mobile application accounts. The only security measures in place rely on SMS or email authentication/verification during account creation (e.g., one-time password and email verification).} In practice, an attacker could rent service from online SMS and Email providers such as Twilio \cite{twilio}, which \textcolor{black}{provide} communication APIs to handle the verification processes. \textcolor{black}{Additionally, attackers can create as many fake email accounts as needed for the verification process.}

\begin{figure}[t]
    \centering
    \includegraphics[width=0.5\linewidth, height=6cm]{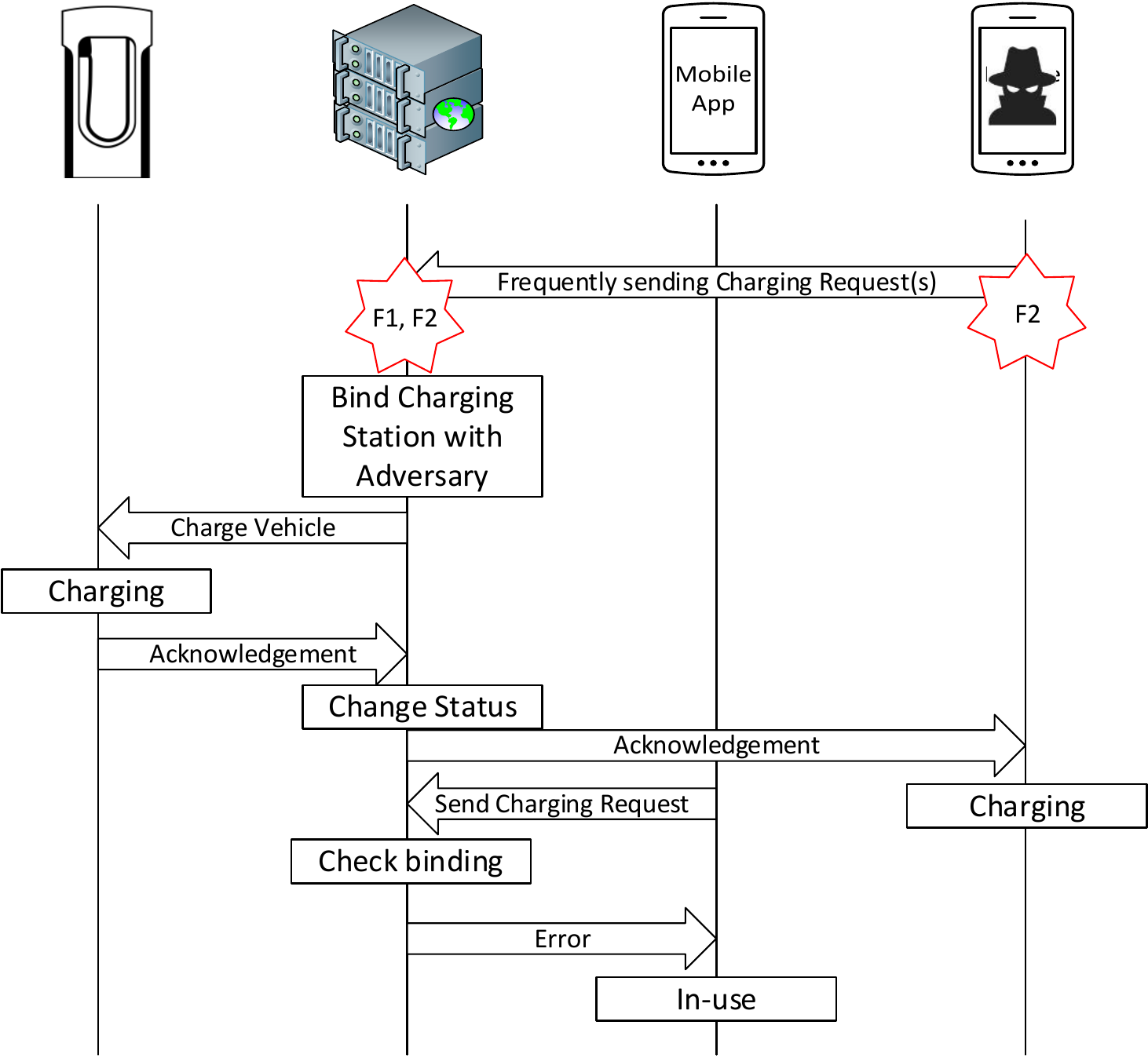}
    \caption{Sequence diagram depicting remote charging/discharging scenario.}
    \label{fig:flaw2}
\end{figure}

\textbf{Remote Charging Session Hijacking.}
\textcolor{black}{After analyzing the interactions of the mobile application with the different entities, we developed an understanding how the mobile application could be used as an attack surface against the power grid.} We found that the studied platforms are vulnerable to session hijacking. \textcolor{black}{By utilizing these vulnerabilities, attackers can initiate unauthorized EV charging sessions with the aim to impact the power grid.} Ideally, the CMS should only allow a charging request if the request is issued from the account owner that is bound with the EV. However, we found that the CMS does not perform any account-based authorization or check during charging. \textcolor{black}{In other words, there is a decoupling between the user account and the EV that is connected to the EVCS.} The user is coupled with the EVCS ID only. Thus, an EV connected to an EVCS can be charged remotely regardless of whether the user initiating the request is the legitimate owner of the EV. \textcolor{black}{The CMS does not perform the necessary checks to validate whether a user is allowed to perform this action,} or if the user is the actual owner of the EV. Thus, allowing adversarial accounts to unlawfully hijack charging sessions.

As illustrated in Figure \ref{fig:flaw2}, the adversary can leverage the combination of F1 and F2 to initiate unauthorized charging requests to take control over the charging session that should have been initiated by the actual EV owner. When the attacker uses an adversarial account to start charging requests, the CMS will establish a connection with the adversary. It is worth mentioning that the actions performed by the adversary are fully legitimate and within the scope of the permitted functionalities of the ecosystem. Consequently, the legitimate EV owners can no longer control their charging session. The only way for the user to stop the charging is by \textcolor{black}{physically unplugging the EV}.

\textcolor{black}{After, the user's charging session has been hijacked, the user can no longer control the EVCS}. While this could raise an alert for a security-savvy user, other users may simply \textcolor{black}{disregard this behavior as long as they see that their EV is charging}. It is worth mentioning that even \textcolor{black}{the security-savvy users will not notice the attackers actions unless they regularly check their mobile applications during the charging process. Additionally, even when the attacked is noticed by these users, the root of the problem cannot be traced back} to the adversary. Only the CMS has the \textcolor{black}{knowledge to trace back the attack's origin. However, due to F1 and F2 described above, the CMS consider the attacker's actions legitimate}.

\textcolor{black}{We describe in Table \ref{table:typesOfEVapps} the possible attack scenarios based on the functionalities instilled in the mobile application showing that 29 out of 31 mobile applications are vulnerable to remote mass charging attacks. We exclude Tata Power EZ Charge and Electrify America as they require inputting the EVCS ID number that is physically placed on the charging station HMI. Moreover, there are only 19 out of the 31 mobile applications that provide remote start and stop of charging and allow adversaries to launch oscillatory load attacks with their advanced control on the stopping of charging.}

\textbf{Remote Discharging.} Vehicle-to-Grid (V2G) capabilities are one of the attractive features of EVs that can one day transform the EV battery into a distributed storage to support the power grid operation. Willing EV owners would register themselves as users willing to contribute \textcolor{black}{to} supporting the grid during peak hours for an incentive (e.g. financial incentives larger than the cost of charging) through utilizing their mobile application. However, as demonstrated above, the current system architecture lacks traceability and end-to-end authentication. An adversary can hijack a session, as explained above, and register themselves as a legitimate user that is willing to contribute \textcolor{black}{to such a V2G scheme}. This allows the adversary to gain monetary compensation by discharging other \textcolor{black}{users'} vehicles. We acknowledge that such \textcolor{black}{a} class of attacks is not feasible at this moment due to the shy adoption of V2G capabilities in the ecosystem and the absence of \textcolor{black}{wide-scale} compensation programs. However, the advancement towards such capabilities being instilled in the ecosystem requires improving the current ecosystem architecture and \textcolor{black}{strict} access control mechanisms, for safe and secure operation.

\subsection{Attack Feasibility}
\label{sec:feasibility}
In this section\textcolor{black}{,} we study the feasibility of launching wide-scale coordinated charging/discharging attacks. In these attacks, we assume that users connect their EVs to the EVCSs before starting a charging session. \textcolor{black}{We also} assume that the vehicle remains connected for a period of time after the end of the charging. In the aims to understand user behavior and predict it, the authors in \cite{chung2019ensemble} highlight that EV owners do not necessarily start charging right after \textcolor{black}{plugging in}. Additionally, a time window exists between plugging in and charging \textcolor{black}{an EV}, which is the time a user needs to pull out the phone to start a charging session. Moreover, according to Almeghrebi et al. \cite{almaghrebi2019analysis}, another time window exists, where customers leave their vehicles for an extended time when parking at the workplace or overnight beyond finishing charging. Some users even leave \textcolor{black}{their} vehicles for longer than 24 hours. This attack window is only applicable \textcolor{black}{to} mobile applications that provide remote start and stop services. These time windows are exploitable by the adversary \textcolor{black}{that} can utilize them to launch remote session hijacking.

In \cite{ma2022multistep}, the authors utilize a multistep hybrid LSTM neural network to predict EVCS occupancy. They base their analysis on public charging data from the City of Dundee, UK in 2018. The number of charging stations \textcolor{black}{plug-in} simultaneously during the day fluctuates reaching 300 charging sessions at 10:00 a.m. during the weekend and 400 charging sessions during the weekday. The number of charging sessions starting at peak times during the day \textcolor{black}{is} expected to increase as more customers adopt electric vehicles as a means of transportation with the rapid and increased deployment of public charging stations. Thus, \textcolor{black}{the} feasibility of remote charging session hijacking at scale increases. 

To execute an attack by exploiting these vulnerabilities, an adversary needs information about the user's behavior. By understanding user behavior, the adversary can time and coordinate the attack to increase its success rate. The attacker can extract information, from the mobile applications (Type 1 and Type 2), similar to \cite{lee2019acn}, where the authors predicted user behavior based on arrival time, duration, departure time, etc. An adversary can utilize the online interface (mobile application/web portals) to gather information. The information we gathered through a tool we devised \textcolor{black}{is the} start charge time (arrival time), \textcolor{black}{and} departure time if a vehicle connected within the attack windows. We used Appium \cite{appium} to automate mobile application scraping which can be used for web applications scraping. We then identify target EVCS and monitor their utilization and status. We collect information about the EVCSs that are in use, allowing us to track arrival and departure times. Moreover, whenever the station's status changes from ``in-use'' to ``available'', we send a probe charging request to check if there is an EV connected to the EVCS. \textcolor{black}{The collected data is used to} model user behavior and allow \textcolor{black}{us and the attacker} to target peak \textcolor{black}{EV connectivity} hours to hijack charging sessions at scale \cite{ma2022multistep}. 

\subsection{Attack Demonstration and Verification}
\begin{figure}
\centering
\begin{minipage}{.5\textwidth}
  \centering
  \includegraphics[width=0.9\linewidth,height=4cm]{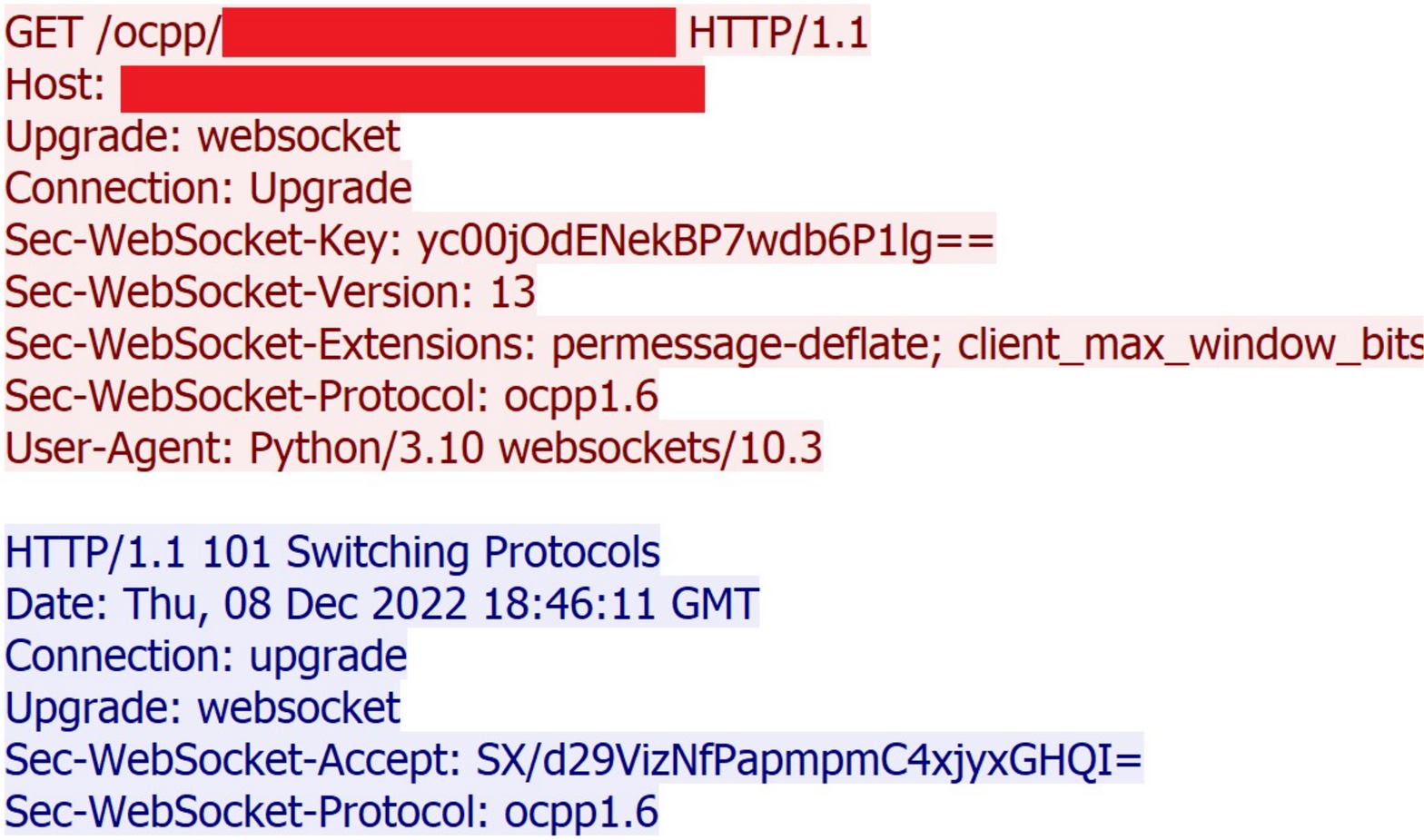}
  \captionof{figure}{EVCS device registration with the CMS using our real-time co-simulation test-bed.}
  \label{fig:httpheader}
\end{minipage}%
\begin{minipage}{.5\textwidth}
  \centering
  \includegraphics[width=0.9\linewidth,height=4cm]{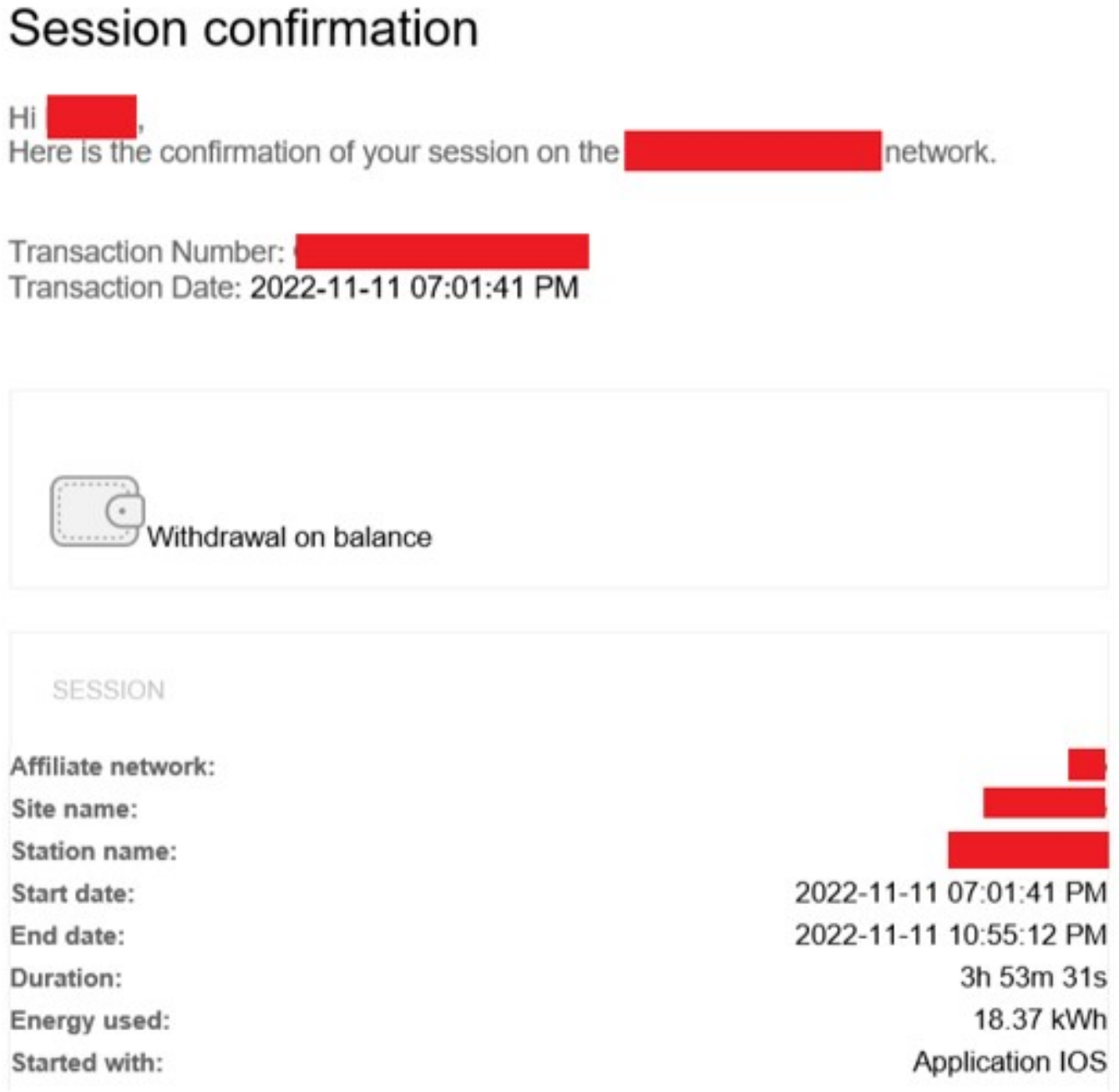}
  \captionof{figure}{Session confirmation showing the success of our attack by hijacking the charging of an idle vehicle.}
  \label{fig:verification}
\end{minipage}
\end{figure}

\textcolor{black}{In this section, we evaluate and verify our observations, inferences, and conclusions by using a real-time co-simulation test bed. We create a replica of the real EVCS ecosystem by integrating real charging station hardware with a production-grade CMS. The EVCS hardware utilizes OCPP v1.6 and communicates with the CMS backend to perform device registration initially. The EVCS and the CMS then continuously communicate with each other over WebSockets to ensure that the EVCS is alive by either sending a heartbeat notification or through the WebSockets ping-pong request/response. Indeed, during device registration, the EVCS will send an HTTP request to the CMS backend which gets upgraded to a WebSocket connection. The HTTP header includes the EVCS identification number which could be the serial number or an operator-defined ID as demonstrated in Figure \ref{fig:httpheader}. Consequently, we were able to confirm and verify our observations and inferences regarding the interactions of the different components discussed previously.}

\textcolor{black}{Accordingly, we aim to demonstrate our new attack vector to verify our conclusions on the vulnerabilities that exist in the EVCS mobile operators. To leverage the aforementioned design flaws that rely on the authentication scheme adopted by the different EVCS mobile operators we first identify two EVCSs mainly in heavily populated areas. We monitor these EVCSs and record their utilization by gathering information from the mobile interface of the applications for 3 days starting the 8$^{th}$ of November 2022 till the 10$^{th}$ of November 2022. During that time we performed reconnaissance to understand the utilization of the EVCS. EVCS$_{1}$ showed heavy arrival during evening hours (between 6:30 PM and 7:30 PM) whereas EVCS$_{2}$ showed heavy arrival between 4:00 PM and 5:00 PM. We note that to identify the utilization and arrival we note the change in the state of the EVCS. When charging an EV, the EVCS shows an ``in-use'' state rendering it unavailable by other users. Consequently, on the 11$^{th}$ of November 2022 we execute our attack as a proof of concept on the EVCSs. We leverage the lack of rate limiting to send multiple charging requests every 3-4 minutes from the adversarial account we created using the legitimate mobile application channels. After sustaining the attack for almost 30 minutes starting at 6:30 PM for EVCS$_{1}$ and starting at 4:00 PM for EVCS$_{2}$. Consequently, at 7:01 PM and at 4:18 PM we were able to successfully hijack the charging session of the EVCS$_{1}$ and EVCS$_{2}$ respectively. Consequently, we show in Figure \ref{fig:verification} the confirmation of a successful charging of a vehicle that does not belong to us for almost 3 hours. Moreover, we would like to note that the attack was demonstrated and verified on two different mobile applications, i.e., one that only allows a start charge functionality and another that allows remote start and stops charging functionality. We also note that vulnerabilities that allow the adversary to remotely start and stop a session could be used in combination with other High-wattage IoTs to impact the power grid. In \cite{soltan2018blackiot}, the authors demonstrate the impact of controlling a large botnet swarm of high-wattage IoTs that could be used to impact the power grid by launching load-altering attacks. Load-altering attacks impact the power grid by inducing grid instability (e.g., frequency instability). Finally, the same attack workflow could be launched to impact the power grid using the other mobile applications as they follow the same procedures and policies to authenticate users.}

\section{Attack Implications}
\label{sec:implications}
In what follows, we discuss the remote charging session hijacking attack scenario along with its implications on the power grid and EV users, respectively.

\subsection{Attack Implications on the Power Grid}
\label{sec:impact-grid}
As demonstrated in the analysis, attackers can leverage the identified vulnerabilities to compromise user accounts and perform synchronized large-scale cyber-attacks against the integrated infrastructure \cite{tony_paper, acharya2020cybersecurity,fraiji2018cyber, pratt2019vehicle}. With the EV charging ecosystem being a new and wide attack surface, it is an attractive target for exploitation by organizations with enough resources to conduct \textcolor{black}{large-scale} attacks against the power grid, by utilizing the mobile application to perform stealthy attacks.

Consequently, an adversary could initiate a distributed botnet attack utilizing thousands of malicious accounts to send charging requests and hijack as many sessions as possible to amplify the attack. The behavior of arrival and departure at charging stations almost coincides with the demand behavior of the power grid \cite{lee2019acn}, as demonstrated in the utility demand curve of California, United States of America \cite{california} and New South Wales (NSW), Australia \cite{aemo_2021}. Additionally, by exploiting the identified vulnerabilities and initiating the remote charging session hijacking attack (Section \ref{sec:attackScenario}) during the described attack windows (Section \ref{sec:feasibility}), an adversary can remotely orchestrate hijacked charging sessions to synchronize a wide scale attack that can disrupt the power grid operations.

In this section, we study the impact of synchronized mass charging attacks on power system economics (i.e., generation cost and transmission line losses). We then examine how adversaries with some knowledge of the grid topology can craft targeted mass charging attacks in order to overload and trip transmission lines. Finally, we study the power grid stability subject to oscillatory load attacks that can cause violation of the safe frequency operation limits and load shedding. \textcolor{black}{Oscillatory load attacks can be performed using 16 of the applications that provide \textcolor{black}{on-and-off} remote control capabilities without requiring the user to scan a QR code.}

To amplify the attack impact on the grid, an adversary with knowledge of the grid can craft targeted and smarter attacks. A small number of compromised charging sessions with enough knowledge of weak buses allow the adversary to disrupt the power grid operations. Power grid information can be estimated through monitoring the measurements of the power grid to estimate the topology, using MILP programming, machine learning\textcolor{black}{,} and voltage and load monitoring \cite{mohammad_paper, cavraro2019real, taheri2019new, cavraro2019inverter, moffat2019unsupervised, gandluru2019joint, deka2019topology, moffat2020real}. Various stability techniques and strategies could then be used by adversaries to locate the most sensitive/vulnerable buses, such as PV and QV curves \cite{mohammad_paper, kundur2007power}.

\begin{figure}[t]
	\centering
	\begin{subfigure}[b]{0.48\textwidth}
		\centering
		\includegraphics[width=\textwidth]{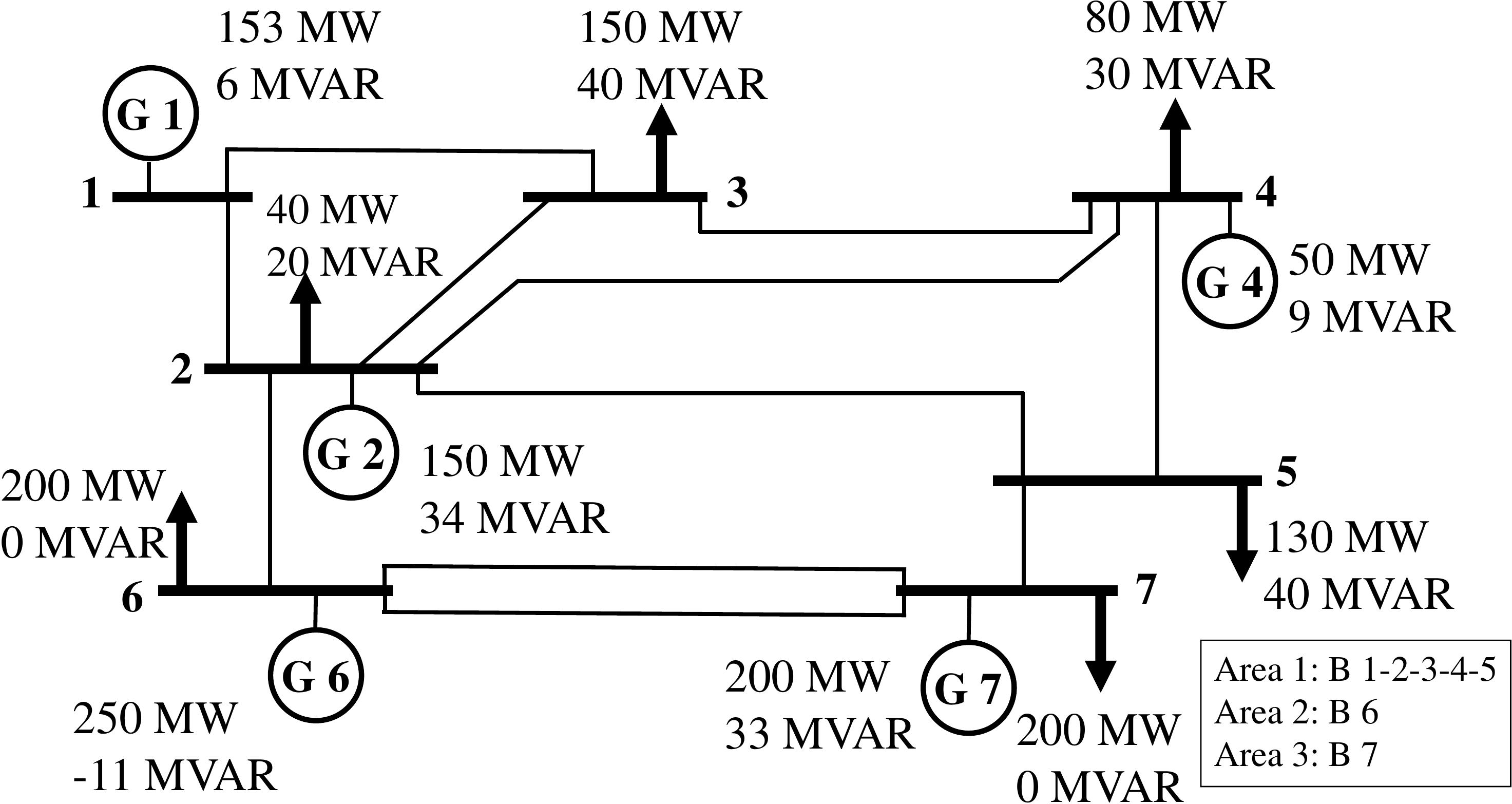}
		\caption{Glover book 7-bus grid}
		\label{fig:grid}
	\end{subfigure}
	\hfill
	\begin{subfigure}[b]{0.48\textwidth}
		\centering
		\includegraphics[width=\textwidth]{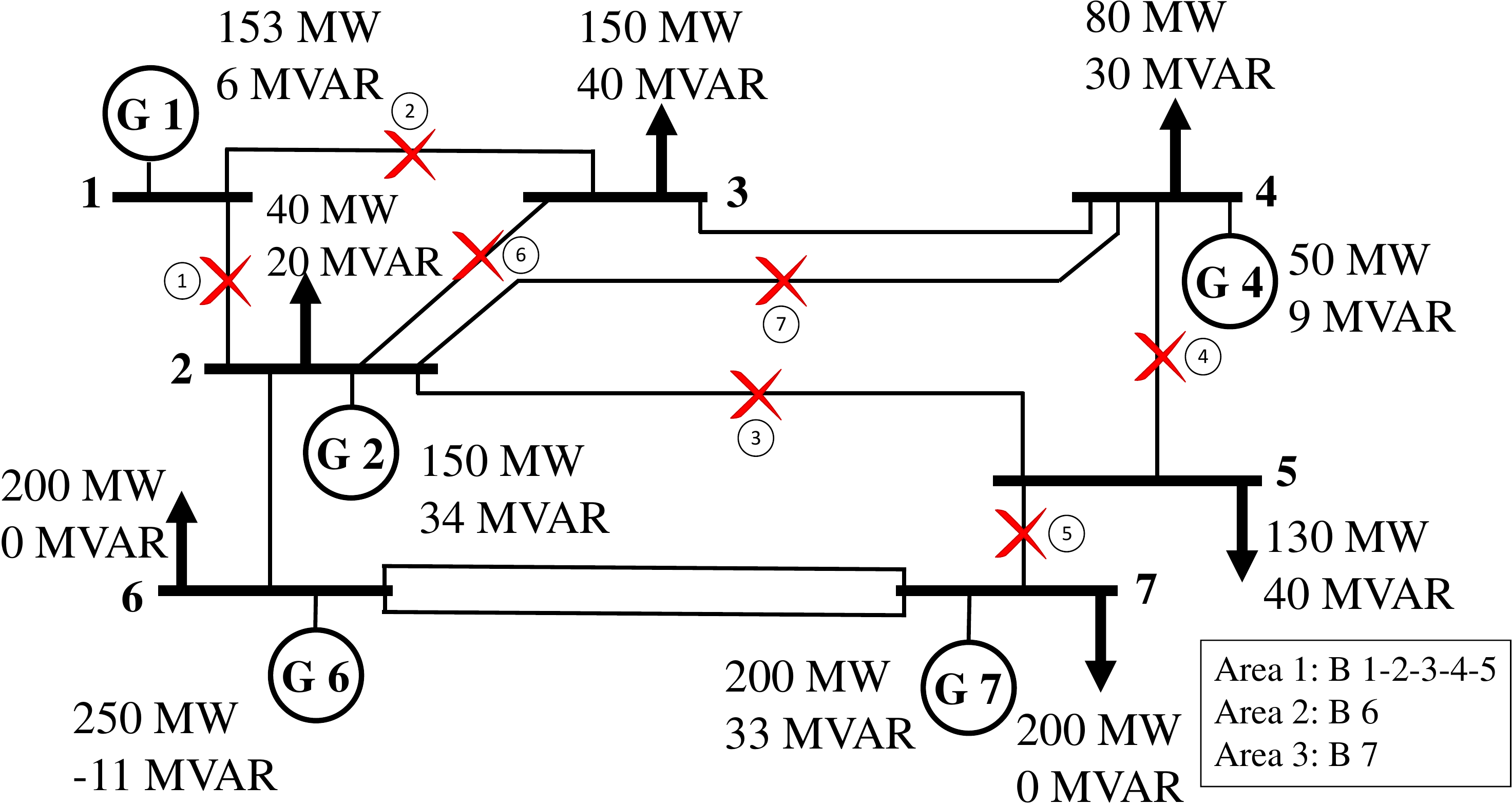}
		\caption{Line tripping attack impact}
		\label{fig:trippedLines}
	\end{subfigure}
	\caption{Overview of the (a) Glover book 7-bus grid and (b) the impact of the line tripping attack scenario.}
\end{figure}

We demonstrate the impact of the attacks on the 7-bus test case introduced by Glover et al. \cite{glover2012power} (Figure \ref{fig:grid}), which is commonly used for research purposes~\cite{mohammad_paper}.
We utilize this grid due to its \textcolor{black}{built-in} optimal power dispatching capabilities, unlike the work in \cite{tony_paper}. Moreover, this 7-bus test case provides the generation costs formulas that will allow us to study the \textcolor{black}{economic} impacts on the utility. To achieve \textcolor{black}{a} close to realistic simulation of the power grid behavior during peak and off-peak demand hours, we scaled the grid loads based on the NSW \cite{aemo_2021} power grid load profile using Equation \ref{eq:equation1}.

\[HourlyLoad_{Scaled} = \frac{GloverLoad \times HourlyLoad_{NSW}}{AverageLoad_{NSW}} \label{eq:equation1} \tag{1}\]

To this end, we use PowerWorld \cite{powerworld_2021} which is a power simulator that allows us to analyze the steady-state power flow, transient stability, generation costs\textcolor{black}{,} and other power system operations. Unlike \cite{mohammad_paper, soltan2018blackiot}, here we study the economical aspects of an attack such as generation cost and line losses respectively. Along with that, we also take into account \textcolor{black}{the} load shedding mechanism that is used by the utility to regulate power generation in case of a sudden drop in frequency below certain thresholds \cite{huang2019not} to demonstrate more realistic attack implications. The different attack simulations and results are demonstrated below. In what follows, the attack is initiated by compromising 84 MW of EV load that \textcolor{black}{is} equivalent to 7636 EVs charging at the 11 kW Level 2 chargers. The current numbers of EVCSs and EVs \textcolor{black}{are} not enough to mount such attacks, however, the growth in the EV numbers will soon provide a large enough surface to make it possible \cite{mohammad_paper}. It is worth noting that mobile applications allow cross-product communication and control, thus, increasing the scale of the attack as more vendors join these platforms. Moreover, as the EVCS market move towards wide adoption of level 3 chargers, the higher power entails higher risk. 

\textbf{Economical impact.}
The attacker can cause the power utility to incur economic losses by launching EV attacks against the power grid. To study the \textcolor{black}{economic} impacts of a mass charging attack on the grid, we examine the transmission line losses and the power generation cost during different loading conditions and under different attack scenarios. \textcolor{black}{To perform mass-charging attacks 30 applications allow us to perform such an attack, whereas the rest prevent remote \textcolor{black}{mass-charging} by forcing the adversary to scan a QR.} We used the scaled load profile to demonstrate the incurred cost and losses at different grid loading conditions. Namely, we focused on the peak load (943 MW), the average load (800 MW)\textcolor{black}{,} and the minimum load (677 MW) conditions that we will refer to later as off-peak load. We simulate 3 different attack scenarios against the test grid by (1) distributing the attack load randomly, (2) distributing the attack load equally among the 6 load buses and (3) distributing the attack load proportionally among the different load buses. It is worth highlighting that Scenario (1) represents a random distribution of the EV charging attack load to simulate an adversary with no knowledge of the grid topology. \textcolor{black}{Scenarios} (2) and (3) represent attacks by an adversary with limited knowledge of the grid and geographical knowledge of load size and EV distribution respectively. 

\begin{figure}[t]
	\centering
	\begin{subfigure}[b]{0.48\textwidth}
		\centering
		\includegraphics[width=\textwidth]{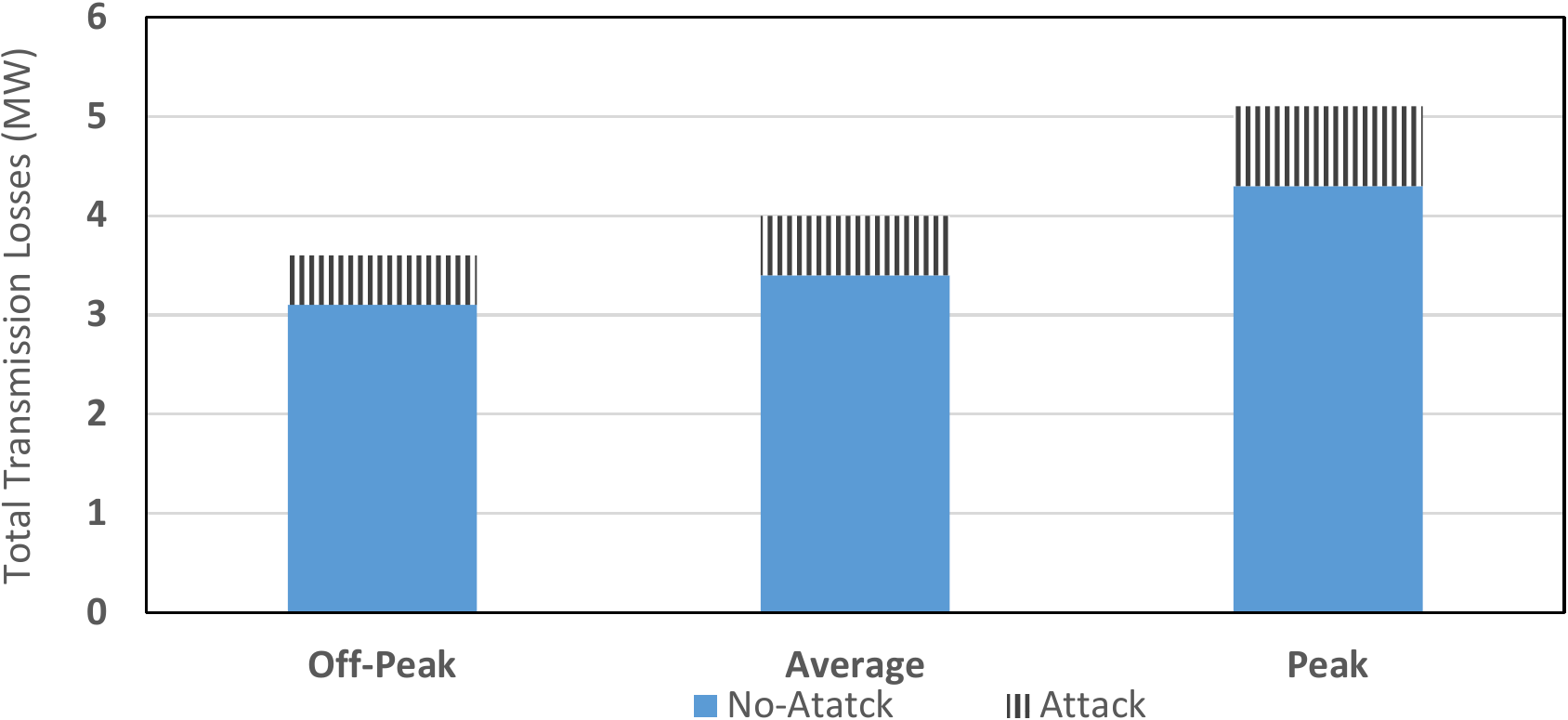}
		\caption{Transmission losses}
		\label{fig:losses}
	\end{subfigure}
	\hfill
	\begin{subfigure}[b]{0.48\textwidth}
		\centering
		\includegraphics[width=\textwidth]{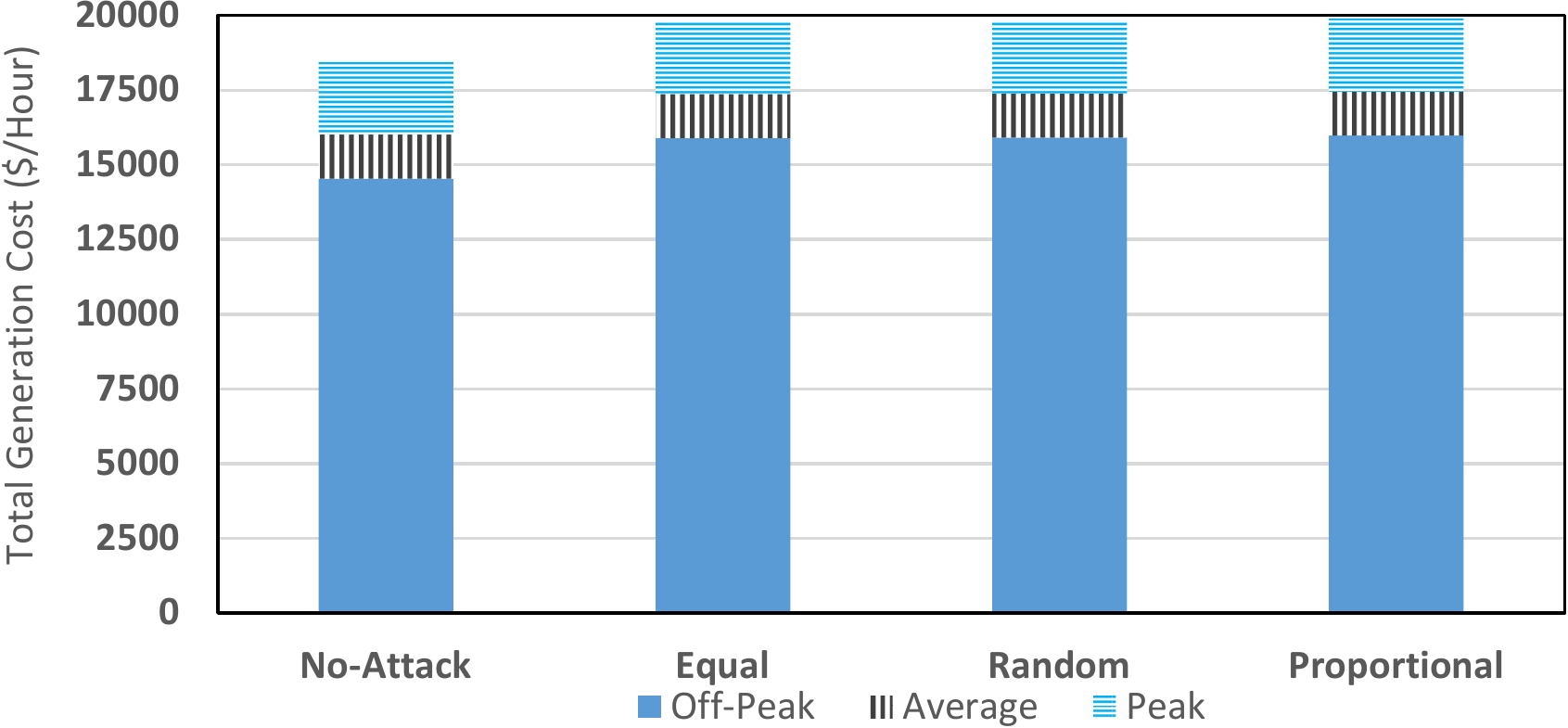}
		\caption{Attack-related costs}
		\label{fig:costs}
	\end{subfigure}
	\caption{Incurred (a) transmission losses and (b) costs due to various attack scenarios.}
	\label{fig:interactions1}
\end{figure}

Generally speaking, the attacks will increase the transmission losses under all loading conditions. However, under higher loading conditions, the same attack will cause more incremental losses due to the increased power flow in the lines. Line losses are calculated as $P_{loss} = R_{line} \times I^{2}$ thus at higher loading conditions, the same attack load will result in more losses. It is worth noting that under the different attack scenarios, the total incremental line losses were almost equal. This is due to the fact that, the total load of the different attack scenarios is the same and that we do not have any extra long transmission lines that will have significantly different losses under different attack load \textcolor{black}{distributions}. The normal and incremental losses are demonstrated in Figure \ref{fig:losses}. The no-attack losses under the different loading conditions were 3.1 MW (off-peak), 3.4 MW (average)\textcolor{black}{,} and 4.3 MW (peak), which lead to an increase \textcolor{black}{of} 16.13\% at off-peak loading conditions, 17.65\% during average loading conditions and 18.6\% during peak conditions. Thus, this simulation clearly demonstrates that the attack impact on system losses is amplified when the power demand was the highest, which also coincides with the time during which EV connection to the chargers is the highest.

In the case of attacks against generation cost, each attack scenario differs based on the optimal power dispatch performed by the utility to \textcolor{black}{reduce} the overall cost. Figure \ref{fig:costs} presents the total generation cost of the system when \textcolor{black}{no attack} occurs and during the three attack scenarios mentioned above. As Figure \ref{fig:costs} demonstrates, the total added cost due to the attack is \textcolor{black}{higher} during peak loading \textcolor{black}{conditions} across all attacks. More importantly, the proportional attack scenario caused the highest extra cost. To put things into context, the no-attack cost at off-peak, average\textcolor{black}{,} and peak loading conditions were \$14,545.28/Hour, \$16,009.39/Hour\textcolor{black}{,} and \$18,438.10/Hour respectively. The added cost due to the proportional attack is \$1,423.83/Hour under off-peak conditions, \$1,426.45/Hour under average conditions\textcolor{black}{,} and \$1,451.95/Hour under peak conditions. This demonstrates how an attacker can force the utility to increase its generation and incur extra costs.

One aspect not present in the simulation was the usage of peak generation units. This was left out due to the absence of these units in the Glover grid in Figure \ref{fig:grid}. These units are usually \textcolor{black}{fast-ramping} units used by utilities and power grid operators during peak hours when the large baseline generation units do not have sufficient capacity to supply all the load. These peak generation units are usually operated for a few hours a day only due to their high operation cost. This means that if the attack occurs at a time \textcolor{black}{when} peak generators are being utilized, the extra cost would be higher. Another aspect of repeated \textcolor{black}{long-term} attack worth mentioning, is that mass charging attacks, especially at peak hours, will cause extra transformer loading. This extra loading would reduce its lifetime and would require more frequent maintenance intervals causing extra maintenance costs.

An attacker with a \textcolor{black}{long-term} target of causing the utility to incur extreme losses can repeat the hijacking of charging sessions over long periods of time. For instance\textcolor{black}{,} launching the above attack for one hour during peak times every day for an entire year will create an extra generation cost \textcolor{black}{totaling} \$529,962 for the utility \textcolor{black}{based on the Glover grid and the generators' cost functions.} To put things into a better context, scaling this attack up to the NSW grid will cause \$4,615,967 extra cost for the utility per year. To this end, an attacker might choose to compromise a smaller number of EV charging sessions and \textcolor{black}{choose} different sets of EVs every day to remain stealthier and still cause millions of dollars of losses to the utility in extra generation costs.

\begin{figure}
     \centering
     \begin{subfigure}[b]{0.48\textwidth}
         \centering
         \includegraphics[width=\textwidth]{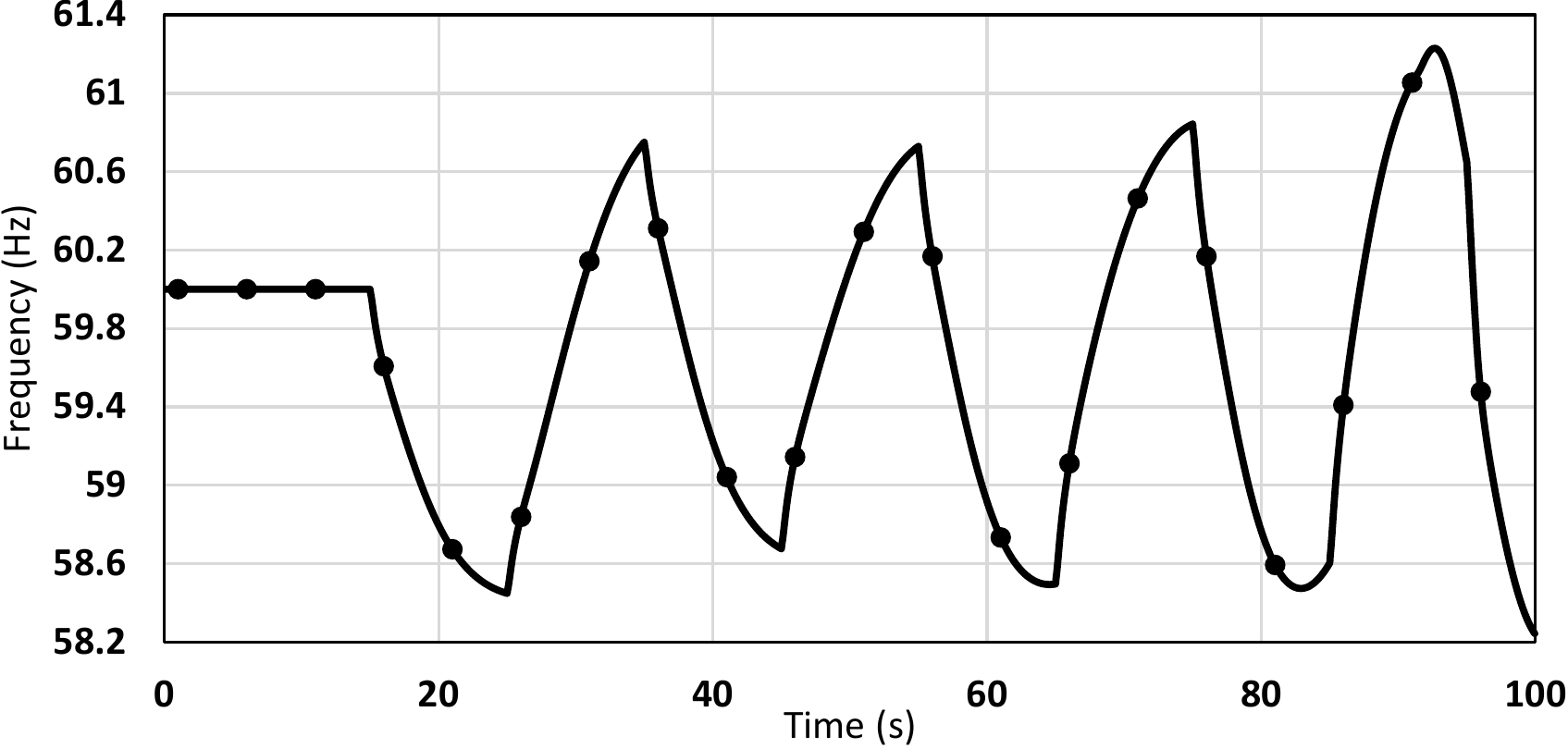}
         \caption{}
         \label{fig:noshedding}
     \end{subfigure}
     \hfill
     \begin{subfigure}[b]{0.48\textwidth}
         \centering
         \includegraphics[width=\textwidth]{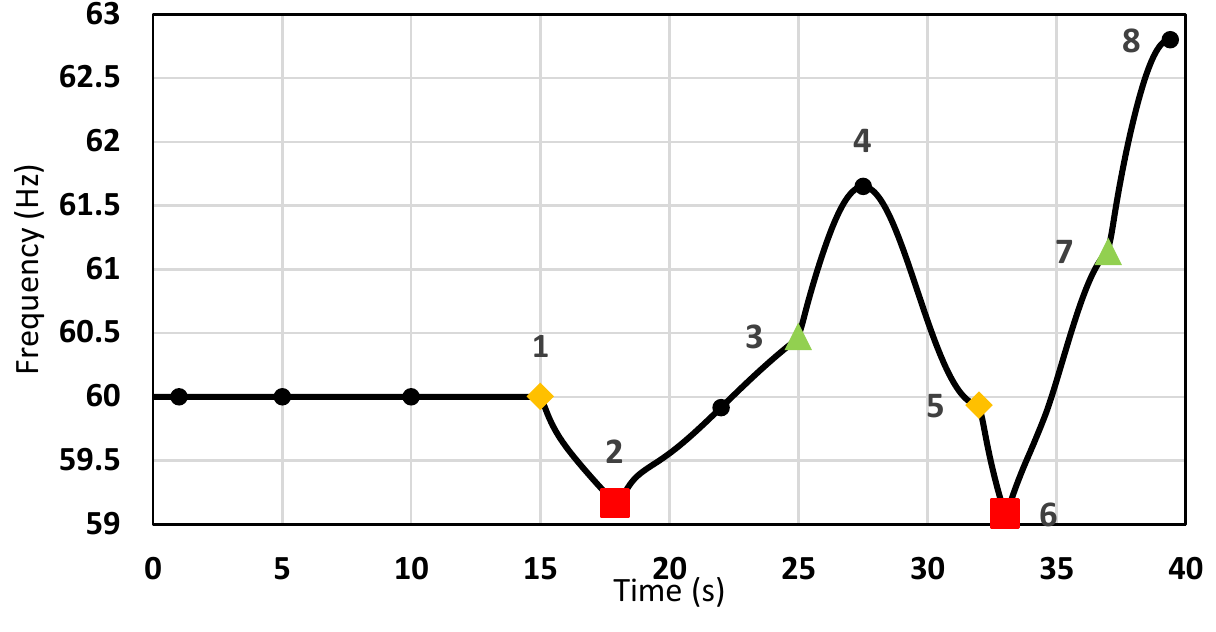}
         \caption{}
          \label{fig:loadshedding}
     \end{subfigure}
        \caption{Frequency behavior over time (a) without load shedding, and (b) with load shedding.}
    \label{fig:interactions2}
\end{figure}

\textbf{Overloading and tripping transmission lines.}
Another type of impact that might be desired by the attacker is causing line overloading and tripping by crafting a targeted attack against the grid. This attack has more severe and immediate consequences since it can leave consumers without electricity. In the previous set of attacks, some lines got highly loaded but none of them reached an overloaded state. The same EV load however can be used by attackers with topology knowledge to target certain lines in order to cause cascading line failure. The attacker will only require knowledge of the topology and estimate values of the loads and power flows but not the line parameters. This information can be found online and in multiple public access databases and websites.

To simulate such attacker behavior, we targeted bus 4 and bus 5 with a \textcolor{black}{synchronized} 20 MW and 64 MW EV charging attack respectively. This attack overloaded and tripped the line connecting bus 1 and bus 2 after which multiple lines would be overloaded and tripped. In total, seven lines would trip successively in the order shown in Figure \ref{fig:trippedLines}. The successive line tripping would lead to islanding each of buses 1, 3, 4\textcolor{black}{,} and 5. While the load at bus 4 will be supplied by power from the generator at the same bus, the loads at bus 3 and 5 will lose their power supply and thus the grid will lose a total of 280 MW which represents a loss of electricity to 35\% of the consumers.

\begin{table*}[t]
    \centering
    \scriptsize
    \caption{Attack Scenario description and impact.}
    \begin{tabular}{c c p{2.5cm} p{2.5cm} c p{4.5cm}}\toprule
                 \textbf{\#}& \textbf{Time(s)} & \textbf{System State} & \textbf{Action} & \textbf{Action By}& \textbf{Impact (state change)}\\\midrule
                1&15&System is operating normally& Total attack load of 40 MW initiated&Attacker& System frequency starts dropping\\\hline
                2&17.9&Frequency drops below 59.3&5\% load shedding& Utility&5\% of total consumers lose electricity. System frequency starts rising\\\hline
                3&25&The frequency peaks and is regulated by the automatic generator control&Turning off all compromised charging sessions&Attacker&System frequency spikes\\\hline
                4&27.5&The frequency starts dropping due to the automatic generation control&Automatic action of generation control system ``no human intervention''&Automatic&System frequency is being reduced to stabilize the system\\\hline
                5&32&The frequency was reduced by the automatic generation control&Total attack load of 80 MW initiated&Attacker&System frequency starts dropping faster than step 1 due to the larger attack load and the reduced generation after load shedding\\\hline
                6&33&Frequency drops below 59.3&5\% load shedding&Utility&Additional 5\% of total consumers lose electricity (10\% total). System frequency starts rising\\\hline
                7&37&The frequency peaks and is regulated by the automatic generator control&Turning off all compromised charging sessions&Attacker&Causes a larger spike in frequency than step 3 since the EV load that was turned off is larger than that of step 3\\\hline
                8&39.4&Frequency exceeds 61.8 Hz \cite{huang2019not}&Generators should be tripped instantaneously&Utility&Sequential generator tripping until system frequency stabilizes.\\\hline
                >8& >39.4&Utility trips generators immediately. The system inertia drops.&The attack impact is larger causing more tripping.&Attacker&As more generators are tripped, the system reachs a state of blackout.\\
        \bottomrule
    \end{tabular}
    \label{table:attack_scenario}
\end{table*}

\textbf{Power grid instability.}
Another attack that takes advantage of load manipulation is an oscillatory load attack that can impact the frequency stability of the power grid. This attack revolves around the concept of creating a demand surge to cause a frequency drop on the grid followed by a drop in demand to cause the frequency to overshoot. In the first step, the attacker will use the compromised accounts and hijack charging sessions to initiate mass charging to increase the power load. This extra power load would create an imbalance between the increased load and the generated power causing the generators to slow down resulting in a frequency drop. The second step of this attack happens when the system starts its recovery. The attacker would switch off the compromised charging sessions, initiated in the first step, to cause a frequency increase that is amplified by the operator’s effort to increase the speed in response to the initial step. The attacker would then alternate between these steps for the desired duration. The impact can be amplified by launching the attack when the system has lower inertia due to the presence of a high share of renewable energy resources.

\textcolor{black}{Given the dependence of the grid’s transient behavior on the generator and turbine models, we utilized the automatic control models common to studies similar to ours. It is important to note that the utilization of different control models will change the exact values of the simulation but the general shape and behavior remain the same. This demonstrates that the attacks can be successful under different conditions, but their magnitudes might \textcolor{black}{need} to be scaled based on the different conditions to achieve the desired impact. In our study, we used the machine model ``GENSAL'', the exciter model ``IEEE T1'' and the turbine governor model ``IEEE G2''.}

The oscillatory load attack is simulated against the grid in Figure \ref{fig:grid} by initiating the oscillatory load behavior described above on buses 3 and 5. The attack is initiated by starting a mass charging session equivalent to 20 MW ( at time t=15s and stopping it at t=25s after the system starts to increase generator speed to compensate and the frequency starts to rise. This charging and stopping behavior \textcolor{black}{are} repeated periodically every 10 seconds while increasing the attack load at each bus by 5.5 MW every cycle. The frequency behavior of the grid that results after such an attack is demonstrated in Figure \ref{fig:noshedding} where we can see the frequency fluctuation due to the oscillatory load behavior. The importance of this attack is that it does not require huge loads to cause the frequency fluctuations depicted in Figure \ref{fig:noshedding}. Even when the compromised EV numbers are much less than \textcolor{black}{in the} example above, a sustained oscillation will hinder the system’s return to normal operation. Sustaining this attack would damage the turbines due to the constant acceleration and deceleration.

In the previous example, we assumed no grid protection mechanism was used by the utility and that the attacker followed a semi-naive approach in which the attack period is predetermined and does not change as a result of actual conditions on the grid. In this iteration of the attack\textcolor{black}{,} we assume the utility will utilize load shedding when the frequency drops below preset thresholds. The threshold that is violated in the attack is the 59.3 Hz threshold after which the utility will immediately disconnect 5\% of the total load in order to compensate for the fast dropping frequency \cite{huang2019not}. This utility behavior is depicted in Figure \ref{fig:loadshedding} by the squares at time t=17.9s and t=33s. \textcolor{black}{The behavior depicted in Figure \ref{fig:loadshedding} is a response to a more advanced oscillatory load attack requiring the attacker to know and observe the actual grid response to tune the attack load and period.} The attacker and utility interaction at every step is summarized in Table \ref{table:attack_scenario}.

An extension of the oscillatory load attack can be achieved by utilizing the reverse power discharge through the V2G functionality similar to a work performed in \cite{kabir2021two}. By initiating this V2G at the instance we stop the mass charging, the attacker can cause a larger frequency spike. It is worth mentioning, that by instilling V2G capabilities in mobile \textcolor{black}{applications} the adversary can then utilize it to increase the oscillatory attack effect on the grid.

\section{Mitigation and Recommended Countermeasures}
\label{sec:countermeasures}
 
\textcolor{black}{The following section provides recommendations for hardening the EVCS ecosystem by reducing the attack surface and addressing the discovered vulnerabilities. First, we provide suggestions based on industry best practices for securing mobile applications. We then provide suggestions based on the unique properties found in the EV ecosystem.}

To assist in mitigating automated attacks, mobile platforms should try to detect bot behavior. To keep malicious software from engaging in abusive behaviors, charging platforms should utilize reCAPTCHA \cite{google} or other similar services which use an advanced risk analysis engine and adaptive challenges. The implementation of reCAPTCHA helps in detecting automated scraping, the creation of synthetic compromised accounts, and automated behaviors. This will not only hinder the attacker from performing the attack, \textcolor{black}{but} it will also hinder the attacker from utilizing mobile applications to perform reconnaissance on the EVCSs and their behavior, which is an integral step in preparation for a wide-scale attack against the grid.

While this could hinder the attacker, it does not prevent the adversary from utilizing the lack of end-to-end authentication. \textcolor{black}{Thus, to mitigate the consequences of such vulnerability we recommend implementing a mechanism for verifying user ownership over the EV. Users should be coupled with their EVs} from the beginning making them the central authority that controls any action performed on their specific EV. Using this we would \textcolor{black}{add} security by design in the interactions between the entities in the EV ecosystem \cite{smartcar}. \textcolor{black}{Each EV has a unique Vehicle Identification Number (VIN) that can be used to identify during the creation of the mobile application account in order to verify that this mobile application user actually has a physical EV. The mobile application should later ask the users to enter the VIN of their EV when they attempt to initiate a charging session. This will ensure that only authorized EVs are charged. This allows the creation of a one-one relationship between the vehicle and the owner}. This would require collaboration between different counterparts such as charging station vendors, and vehicle manufacturers. 

To ensure that the decision-making is distributed \textcolor{black}{and more robust against cyber tampering}, we \textcolor{black}{recommend that the EVCS needs} to check if a connected vehicle has the same VIN sent in the charging request. If the identifiers match, then the EVCS will start charging \textcolor{black}{after verifying the} end-to-end authentication. If no match was found, then it will return an error to the user. The adversary can no longer register to the platform without registering with a valid VIN. \textcolor{black}{Furthermore, even if the adversary creates an account using a legitimate VIN, the adversary will not be able to know the specific VIN of the EV connected at the targeted EVCS}. Thus, by enabling a strict end-to-end authentication the attacker can not be associated with someone else's EV. This would prevent the adversary from launching large-scale attacks against the grid.

\textcolor{black}{Additionally, EV charging can be restricted to users who are in the vicinity of their EVs and the EVCS. Some EVs have NFC chips installed that can be used to initiate charging. The mobile application can use NFC to communicate with the EV and verify the user is in close proximity to their EV. The mobile application can use GPS-based location information to confirm that the user is in proximity to the EVCS. This will ensure that only authorized users who are physically present near the EV and EVCS can initiate charging.} 

\textcolor{black}{We understand that such restrictions might affect the usability and commercialization of the mobile application. For example, when Alice lends her car to her son, he should be allowed to charge the EV. Thus we suggest creating an authorization list where Alice, can add a list of verified mobile application accounts that are authorized to control the charging of the EV}.

\textcolor{black}{By implementing the suggested mitigation methods, the security of the EV charging mobile application will be hardened and the attacks discussed above will be rendered extremely difficult to perform.}

\section{Conclusion}
 \label{sec:conclusion}
In this work, we explored the security of the EV charging ecosystem by focusing on the understudied EV charging mobile application as a main attack surface. We studied the interactions of the mobile application with other components to understand its remote control functionality over the charging stations. Our analysis of the identified interactions and communications unveiled critical vulnerabilities that allow remote adversaries to gain control over the charging operations and perform DDoS attacks by preventing legitimate users from using the charging equipment. 
Moreover, while we demonstrate the feasibility of exploiting existing  EV charging mobile applications' vulnerabilities to hijack charging sessions, we discuss the implications of such attacks on various stakeholders within the EV charging ecosystem. Specifically, we discuss the impact of \textcolor{black}{wide-scale} remote attacks \textcolor{black}{on} the underlying critical infrastructure (i.e., the power grid) and show that an attacker can utilize a botnet of adversarial accounts on those vulnerable mobile applications to cripple the operations of the power grid. Finally, while we discuss attack implications against EV consumers, we also recommend \textcolor{black}{countermeasures} to secure the infrastructure and impede adversaries from performing reconnaissance and launching remote attacks using compromised accounts. In future work, \textcolor{black}{we aim at studying different distributed blockchain solutions to mitigate these design flaws while minimizing integration overhead for the operators}.

\section{Acknowledgement}
\textcolor{black}{This research was conducted and funded as part of the Concordia University/ Hydro-Quebec/ NSERC research collaboration project ``Large-Scale Integration of EVCSs into the Smart Grid: A comprehensive cyber-physical study and security assessment." Grant reference: ALLRP 567144-21.}

\bibliographystyle{IEEEtran}
\bibliography{sample-base}

\begin{thebibliography}{10}
\providecommand{\url}[1]{#1}
\csname url@samestyle\endcsname
\providecommand{\newblock}{\relax}
\providecommand{\bibinfo}[2]{#2}
\providecommand{\BIBentrySTDinterwordspacing}{\spaceskip=0pt\relax}
\providecommand{\BIBentryALTinterwordstretchfactor}{4}
\providecommand{\BIBentryALTinterwordspacing}{\spaceskip=\fontdimen2\font plus
\BIBentryALTinterwordstretchfactor\fontdimen3\font minus
  \fontdimen4\font\relax}
\providecommand{\BIBforeignlanguage}[2]{{%
\expandafter\ifx\csname l@#1\endcsname\relax
\typeout{** WARNING: IEEEtran.bst: No hyphenation pattern has been}%
\typeout{** loaded for the language `#1'. Using the pattern for}%
\typeout{** the default language instead.}%
\else
\language=\csname l@#1\endcsname
\fi
#2}}
\providecommand{\BIBdecl}{\relax}
\BIBdecl

\bibitem{regan_2020}
\BIBentryALTinterwordspacing
H.~Regan, ``China pledges to go carbon neutral by 2060,'' Sep 2020. [Online].
  Available:
  \url{https://www.cnn.com/2020/09/22/china/xi-jinping-carbon-neutral-2060-intl-hnk/index.html}
\BIBentrySTDinterwordspacing

\bibitem{gyulai_2020}
\BIBentryALTinterwordspacing
L.~Gyulai, ``Montreal's climate plan includes ban on non-electric cars downtown
  by 2030,'' Dec 2020. [Online]. Available:
  \url{https://montrealgazette.com/news/local-news/montreal-releases-climate-plan-including-ban-on-non-electric-cars-downtown-by-2030}
\BIBentrySTDinterwordspacing

\bibitem{riley_2021}
\BIBentryALTinterwordspacing
C.~Riley, ``Europe aims to kill gasoline and diesel cars by 2035,'' Aug 2021.
  [Online]. Available:
  \url{https://edition.cnn.com/2021/07/14/business/eu-emissions-climate-cars/index.html}
\BIBentrySTDinterwordspacing

\bibitem{canada_2021}
\BIBentryALTinterwordspacing
N.~R. Canada, ``Government of canada,'' Oct 2021. [Online]. Available:
  \url{https://www.nrcan.gc.ca/energy-efficiency/transportation-alternative-fuels/zero-emission-vehicle-infrastructure-program/21876}
\BIBentrySTDinterwordspacing

\bibitem{acharya2020cybersecurity}
S.~Acharya, Y.~Dvorkin, H.~Pand{\v{z}}i{\'c}, and R.~Karri, ``Cybersecurity of
  smart electric vehicle charging: A power grid perspective,'' \emph{IEEE
  Access}, vol.~8, pp. 214\,434--214\,453, 2020.

\bibitem{mohammad_paper}
\BIBentryALTinterwordspacing
M.~A. Sayed, R.~Atallah, C.~Assi, and M.~Debbabi, ``Electric vehicle attack
  impact on power grid operation,'' \emph{International Journal of Electrical
  Power \& Energy Systems}, p. 107784, 2021. [Online]. Available:
  \url{https://www.sciencedirect.com/science/article/pii/S0142061521010048}
\BIBentrySTDinterwordspacing

\bibitem{tony_paper}
T.~Nasr, S.~Torabi, E.~Bou-Harb, C.~Fachkha, and C.~Assi, ``Power jacking your
  station: In-depth security analysis of electric vehicle charging station
  management systems,'' \emph{Computers \& Security}, p. 102511, 2021.

\bibitem{akhras2020securing}
R.~Akhras, W.~El-Hajj, M.~Majdalani, H.~Hajj, R.~Jabr, and K.~Shaban,
  ``Securing smart grid communication using ethereum smart contracts,'' in
  \emph{2020 International Wireless Communications and Mobile Computing
  (IWCMC)}.\hskip 1em plus 0.5em minus 0.4em\relax IEEE, 2020, pp. 1672--1678.

\bibitem{alcaraz2017ocpp}
C.~Alcaraz, J.~Lopez, and S.~Wolthusen, ``Ocpp protocol: Security threats and
  challenges,'' \emph{IEEE Transactions on Smart Grid}, vol.~8, no.~5, pp.
  2452--2459, 2017.

\bibitem{rubio2018addressing}
J.~E. Rubio, C.~Alcaraz, and J.~Lopez, ``Addressing security in ocpp:
  Protection against man-in-the-middle attacks,'' in \emph{2018 9th IFIP
  International Conference on New Technologies, Mobility and Security
  (NTMS)}.\hskip 1em plus 0.5em minus 0.4em\relax IEEE, 2018, pp. 1--5.

\bibitem{elhussini2021tale}
H.~ElHussini, C.~Assi, B.~Moussa, R.~Atallah, and A.~Ghrayeb, ``A tale of two
  entities: Contextualizing the security of electric vehicle charging stations
  on the power grid,'' \emph{ACM Transactions on Internet of Things}, vol.~2,
  no.~2, pp. 1--21, 2021.

\bibitem{antoun2020detailed}
J.~Antoun, M.~E. Kabir, B.~Moussa, R.~Atallah, and C.~Assi, ``A detailed
  security assessment of the ev charging ecosystem,'' \emph{IEEE Network},
  vol.~34, no.~3, pp. 200--207, 2020.

\bibitem{ocpp}
\BIBentryALTinterwordspacing
``Ocpp 2.0.1, protocols, home,'' 2021. [Online]. Available:
  \url{https://www.openchargealliance.org/protocols/ocpp-201/}
\BIBentrySTDinterwordspacing

\bibitem{baker2019losing}
R.~Baker and I.~Martinovic, ``Losing the car keys: Wireless phy-layer
  insecurity in $\{$EV$\}$ charging,'' in \emph{28th $\{$USENIX$\}$ Security
  Symposium ($\{$USENIX$\}$ Security 19)}, 2019, pp. 407--424.

\bibitem{wong2006battery}
Y.~Wong, K.~Chau, and C.~Chan, ``Battery sizing for plug-in hybrid electric
  vehicles,'' \emph{Journal of Asian Electric Vehicles}, vol.~4, no.~2, pp.
  899--904, 2006.

\bibitem{USDepEnergy}
\BIBentryALTinterwordspacing
``How do fuel cell electric vehicles work using hydrogen?'' 2021. [Online].
  Available:
  \url{https://afdc.energy.gov/vehicles/how-do-fuel-cell-electric-cars-work}
\BIBentrySTDinterwordspacing

\bibitem{kaspersky}
S.~Dmitry, ``{ChargePoint Home Security Research},''
  \url{https://media.kasperskycontenthub.com/wp-content/uploads/sites/43/2018/12/13084354/ChargePoint-Home-security-research_final.pdf},
  Dec. 2018.

\bibitem{occpdef}
J.~E. Rubio, C.~Alcaraz, and J.~Lopez, ``{Addressing Security in OCPP:
  Protection Against Man-in-the-Middle Attacks},'' in \emph{2018 9th IFIP
  International Conference on New Technologies, Mobility and Security (NTMS)},
  Paris, France, 2018, pp. 1--5.

\bibitem{khan2019impact}
O.~G.~M. Khan, E.~El-Saadany, A.~Youssef, and M.~Shaaban, ``Impact of electric
  vehicles botnets on the power grid,'' in \emph{2019 IEEE Electrical Power and
  Energy Conference (EPEC)}.\hskip 1em plus 0.5em minus 0.4em\relax IEEE, 2019,
  pp. 1--5.

\bibitem{clement2009impact}
K.~Clement-Nyns, E.~Haesen, and J.~Driesen, ``The impact of charging plug-in
  hybrid electric vehicles on a residential distribution grid,'' \emph{IEEE
  Transactions on power systems}, vol.~25, no.~1, pp. 371--380, 2009.

\bibitem{leemput2014impact}
N.~Leemput, F.~Geth, J.~Van~Roy, A.~Delnooz, J.~B{\"u}scher, and J.~Driesen,
  ``Impact of electric vehicle on-board single-phase charging strategies on a
  flemish residential grid,'' \emph{IEEE Transactions on Smart Grid}, vol.~5,
  no.~4, pp. 1815--1822, 2014.

\bibitem{dubey2015electric}
A.~Dubey and S.~Santoso, ``Electric vehicle charging on residential
  distribution systems: Impacts and mitigations,'' \emph{IEEE Access}, vol.~3,
  pp. 1871--1893, 2015.

\bibitem{morais2014evaluation}
H.~Morais, T.~Sousa, Z.~Vale, and P.~Faria, ``Evaluation of the electric
  vehicle impact in the power demand curve in a smart grid environment,''
  \emph{Energy Conversion and Management}, vol.~82, pp. 268--282, 2014.

\bibitem{shafiee2013investigating}
S.~Shafiee, M.~Fotuhi-Firuzabad, and M.~Rastegar, ``Investigating the impacts
  of plug-in hybrid electric vehicles on power distribution systems,''
  \emph{IEEE Transactions on Smart Grid}, vol.~4, no.~3, pp. 1351--1360, 2013.

\bibitem{soykan2021disrupting}
E.~U. Soykan, M.~Bagriyanik, and G.~Soykan, ``Disrupting the power grid via ev
  charging: The impact of the sms phishing attacks,'' \emph{Sustainable Energy,
  Grids and Networks}, vol.~26, p. 100477, 2021.

\bibitem{zhou2019discovering}
W.~Zhou, Y.~Jia, Y.~Yao, L.~Zhu, L.~Guan, Y.~Mao, P.~Liu, and Y.~Zhang,
  ``Discovering and understanding the security hazards in the interactions
  between iot devices, mobile apps, and clouds on smart home platforms,'' in
  \emph{28th $\{$USENIX$\}$ Security Symposium ($\{$USENIX$\}$ Security 19)},
  2019, pp. 1133--1150.

\bibitem{dea_29_2021}
\BIBentryALTinterwordspacing
P.~by~S.~O'Dea and J.~29, ``Mobile os market share 2021,'' Jun 2021. [Online].
  Available:
  \url{https://www.statista.com/statistics/272698/global-market-share-held-by-mobile-operating-systems-since-2009/}
\BIBentrySTDinterwordspacing

\bibitem{apktool}
\BIBentryALTinterwordspacing
``apktool - a tool for reverse engineering 3rd party, closed, binary android
  apps.'' 2021. [Online]. Available:
  \url{https://ibotpeaches.github.io/Apktool/}
\BIBentrySTDinterwordspacing

\bibitem{dex2jar}
\BIBentryALTinterwordspacing
``Dex2jar: Kali linux tools,'' Oct 2021. [Online]. Available:
  \url{https://www.kali.org/tools/dex2jar/}
\BIBentrySTDinterwordspacing

\bibitem{jdgui}
\BIBentryALTinterwordspacing
``Jd-gui: Kali linux tools,'' Oct 2021. [Online]. Available:
  \url{https://www.kali.org/tools/jd-gui/}
\BIBentrySTDinterwordspacing

\bibitem{mobsf}
\BIBentryALTinterwordspacing
MobSF, ``Mobile security framework (mobsf),'' 2021. [Online]. Available:
  \url{https://github.com/MobSF/Mobile-Security-Framework-MobSF}
\BIBentrySTDinterwordspacing

\bibitem{literadar}
\BIBentryALTinterwordspacing
Pkumza, ``pkumza/literadar: Lite version of libradar,'' 2021. [Online].
  Available: \url{https://github.com/pkumza/LiteRadar}
\BIBentrySTDinterwordspacing

\bibitem{yang2015static}
S.~Yang, D.~Yan, H.~Wu, Y.~Wang, and A.~Rountev, ``Static control-flow analysis
  of user-driven callbacks in android applications,'' in \emph{2015 IEEE/ACM
  37th IEEE International Conference on Software Engineering}, vol.~1.\hskip
  1em plus 0.5em minus 0.4em\relax IEEE, 2015, pp. 89--99.

\bibitem{azim2013targeted}
T.~Azim and I.~Neamtiu, ``Targeted and depth-first exploration for systematic
  testing of android apps,'' in \emph{Proceedings of the 2013 ACM SIGPLAN
  international conference on Object oriented programming systems languages \&
  applications}, 2013, pp. 641--660.

\bibitem{spoofLocation}
\BIBentryALTinterwordspacing
``Gps joystick guide – the app ninjas.'' [Online]. Available:
  \url{http://gpsjoystick.theappninjas.com/}
\BIBentrySTDinterwordspacing

\bibitem{says_2021}
\BIBentryALTinterwordspacing
S.~says, ``Virtualxposed apk 0.20.3 download latest in 2021 [official],'' Dec
  2021. [Online]. Available: \url{https://virtualxposed.com/}
\BIBentrySTDinterwordspacing

\bibitem{ac-pm}
\BIBentryALTinterwordspacing
Ac-Pm, ``Ac-pm/inspeckage: Android package inspector - dynamic analysis with
  api hooks, start unexported activities and more. (xposed module).'' [Online].
  Available: \url{https://github.com/ac-pm/Inspeckage}
\BIBentrySTDinterwordspacing

\bibitem{portswigger}
\BIBentryALTinterwordspacing
``Burp suite - application security testing software,'' 2021. [Online].
  Available: \url{https://portswigger.net/burp}
\BIBentrySTDinterwordspacing

\bibitem{schmutzler2013evaluation}
J.~Schmutzler, C.~A. Andersen, and C.~Wietfeld, ``Evaluation of ocpp and iec
  61850 for smart charging electric vehicles,'' \emph{World Electric Vehicle
  Journal}, vol.~6, no.~4, pp. 863--874, 2013.

\bibitem{schmutzler2012distributed}
J.~Schmutzler, C.~Wietfeld, and C.~A. Andersen, ``Distributed energy resource
  management for electric vehicles using iec 61850 and iso/iec 15118,'' in
  \emph{2012 IEEE Vehicle Power and Propulsion Conference}.\hskip 1em plus
  0.5em minus 0.4em\relax IEEE, 2012, pp. 1457--1462.

\bibitem{twilio}
\BIBentryALTinterwordspacing
``Communication apis for sms, voice, video \& authentication.'' [Online].
  Available: \url{https://www.twilio.com/}
\BIBentrySTDinterwordspacing

\bibitem{chung2019ensemble}
Y.-W. Chung, B.~Khaki, T.~Li, C.~Chu, and R.~Gadh, ``Ensemble machine
  learning-based algorithm for electric vehicle user behavior prediction,''
  \emph{Applied Energy}, vol. 254, p. 113732, 2019.

\bibitem{almaghrebi2019analysis}
A.~Almaghrebi, S.~Shom, F.~Al~Juheshi, K.~James, and M.~Alahmad, ``Analysis of
  user charging behavior at public charging stations,'' in \emph{2019 IEEE
  Transportation Electrification Conference and Expo (ITEC)}.\hskip 1em plus
  0.5em minus 0.4em\relax IEEE, 2019, pp. 1--6.

\bibitem{ma2022multistep}
T.-Y. Ma and S.~Faye, ``Multistep electric vehicle charging station occupancy
  prediction using hybrid lstm neural networks,'' \emph{Energy}, p. 123217,
  2022.

\bibitem{lee2019acn}
Z.~J. Lee, T.~Li, and S.~H. Low, ``Acn-data: Analysis and applications of an
  open ev charging dataset,'' in \emph{Proceedings of the Tenth ACM
  International Conference on Future Energy Systems}, 2019, pp. 139--149.

\bibitem{appium}
\BIBentryALTinterwordspacing
``Automation for apps,'' 2021. [Online]. Available: \url{https://appium.io/}
\BIBentrySTDinterwordspacing

\bibitem{soltan2018blackiot}
S.~Soltan, P.~Mittal, and H.~V. Poor, ``Blackiot: Iot botnet of high wattage
  devices can disrupt the power grid,'' in \emph{27th $\{$USENIX$\}$ Security
  Symposium ($\{$USENIX$\}$ Security 18)}, 2018, pp. 15--32.

\bibitem{fraiji2018cyber}
Y.~Fraiji, L.~B. Azzouz, W.~Trojet, and L.~A. Saidane, ``Cyber security issues
  of internet of electric vehicles,'' in \emph{2018 IEEE Wireless
  Communications and Networking Conference (WCNC)}.\hskip 1em plus 0.5em minus
  0.4em\relax IEEE, 2018, pp. 1--6.

\bibitem{pratt2019vehicle}
R.~M. Pratt and T.~E. Carroll, ``Vehicle charging infrastructure security,'' in
  \emph{2019 IEEE International Conference on Consumer Electronics
  (ICCE)}.\hskip 1em plus 0.5em minus 0.4em\relax IEEE, 2019, pp. 1--5.

\bibitem{california}
\BIBentryALTinterwordspacing
``California iso - demand trend,'' 2021. [Online]. Available:
  \url{https://www.caiso.com/TodaysOutlook/Pages/default.aspx}
\BIBentrySTDinterwordspacing

\bibitem{aemo_2021}
\BIBentryALTinterwordspacing
``Australian energy - demand trend,'' Oct 2021. [Online]. Available:
  \url{https://aemo.com.au/en}
\BIBentrySTDinterwordspacing

\bibitem{cavraro2019real}
G.~Cavraro, A.~Bernstein, V.~Kekatos, and Y.~Zhang, ``Real-time identifiability
  of power distribution network topologies with limited monitoring,''
  \emph{IEEE Control Systems Letters}, vol.~4, no.~2, pp. 325--330, 2019.

\bibitem{taheri2019new}
S.~I. Taheri, M.~Salles, and N.~Kagan, ``A new modified tlbo algorithm for
  placement of avrs in distribution system,'' in \emph{2019 IEEE PES Innovative
  Smart Grid Technologies Conference-Latin America (ISGT Latin America)}.\hskip
  1em plus 0.5em minus 0.4em\relax IEEE, 2019, pp. 1--6.

\bibitem{cavraro2019inverter}
G.~Cavraro and V.~Kekatos, ``Inverter probing for power distribution network
  topology processing,'' \emph{IEEE Transactions on Control of Network
  Systems}, vol.~6, no.~3, pp. 980--992, 2019.

\bibitem{moffat2019unsupervised}
K.~Moffat, M.~Bariya, and A.~Von~Meier, ``Unsupervised impedance and topology
  estimation of distribution networks—limitations and tools,'' \emph{IEEE
  Transactions on Smart Grid}, vol.~11, no.~1, pp. 846--856, 2019.

\bibitem{gandluru2019joint}
A.~Gandluru, S.~Poudel, and A.~Dubey, ``Joint estimation of operational
  topology and outages for unbalanced power distribution systems,'' \emph{IEEE
  Transactions on Power Systems}, vol.~35, no.~1, pp. 605--617, 2019.

\bibitem{deka2019topology}
D.~Deka, M.~Chertkov, and S.~Backhaus, ``Topology estimation using graphical
  models in multi-phase power distribution grids,'' \emph{IEEE Transactions on
  Power Systems}, vol.~35, no.~3, pp. 1663--1673, 2019.

\bibitem{moffat2020real}
K.~Moffat, M.~Bariya, and A.~Von~Meier, ``Real time effective impedance
  estimation for power system state estimation,'' in \emph{2020 IEEE Power \&
  Energy Society Innovative Smart Grid Technologies Conference (ISGT)}.\hskip
  1em plus 0.5em minus 0.4em\relax IEEE, 2020, pp. 1--5.

\bibitem{kundur2007power}
P.~Kundur, ``Power system stability,'' \emph{Power system stability and
  control}, pp. 7--1, 2007.

\bibitem{glover2012power}
J.~D. Glover, M.~S. Sarma, and T.~Overbye, \emph{Power system analysis \&
  design, SI version}.\hskip 1em plus 0.5em minus 0.4em\relax Cengage Learning,
  2012.

\bibitem{powerworld_2021}
\BIBentryALTinterwordspacing
``Powerworldthe visual approach to electric power systems,'' Jun 2021.
  [Online]. Available: \url{https://www.powerworld.com/}
\BIBentrySTDinterwordspacing

\bibitem{huang2019not}
B.~Huang, A.~A. Cardenas, and R.~Baldick, ``Not everything is dark and gloomy:
  Power grid protections against iot demand attacks,'' in \emph{28th
  $\{$USENIX$\}$ Security Symposium ($\{$USENIX$\}$ Security 19)}, 2019, pp.
  1115--1132.

\bibitem{kabir2021two}
M.~E. Kabir, M.~Ghafouri, B.~Moussa, and C.~Assi, ``A two-stage protection
  method for detection and mitigation of coordinated evse switching attacks,''
  \emph{IEEE Transactions on Smart Grid}, 2021.

\bibitem{google}
\BIBentryALTinterwordspacing
``Recaptcha,'' 2021. [Online]. Available:
  \url{https://www.google.com/recaptcha/about/}
\BIBentrySTDinterwordspacing

\bibitem{smartcar}
\BIBentryALTinterwordspacing
``Smartcar · api platform for connected car data.'' [Online]. Available:
  \url{https://smartcar.com/}
\BIBentrySTDinterwordspacing

\end{thebibliography}
\end{document}